\documentclass[aps, prper, reprint, superscriptaddress, showpacs, nofootinbib, longbibliography, floatfix, raggedbottom]{revtex4-2}

\usepackage[T1]{fontenc}
\usepackage{mathptmx}
\usepackage{array}[=2016-10-06]
\usepackage{longtable}[=2020-01-02]
\usepackage{ragged2e}
\usepackage{needspace}
\usepackage{tabularx} 

\usepackage[utf8]{inputenc}
\usepackage[english]{babel}
\usepackage{booktabs}
\usepackage{amsmath,amssymb,bm}
\usepackage[dvipsnames]{xcolor}
\usepackage{graphicx}
\usepackage{hyperref}
\usepackage[margin=0.75in]{geometry}

\hypersetup{
  colorlinks=true,
  hypertexnames=false,
  linktocpage, 
  urlcolor=magenta!90!black,    
  linkcolor=blue!60!black, 
  citecolor=black!60 
}

\makeatletter
\def\l@subsubsection#1#2{}
\makeatother

\begin{document}

\title{\texorpdfstring{Addressing misconceptions in university physics:\\A review and experiences from quantum physics educators}{Addressing misconceptions in university physics: A review and experiences from quantum physics educators}}

\author{Shayan Majidy}
\email{smajidy@fas.harvard.edu}
\affiliation{Department of Physics, Harvard University, Cambridge, Massachusetts 02138, USA}

\date{September 21, 2026}

\begin{abstract}
Students often begin physics courses with misconceptions rooted in everyday experience and intuition, which can be resistant to change. While research has identified strategies for addressing misconceptions across physics, it remains unclear whether different domains, like classical and quantum physics, require different approaches. Quantum concepts can pose different representational and mathematical demands, while both domains require instructors to elicit and develop students' reasoning. To support discussion of these instructional challenges, we organize selected literature on addressing misconceptions in physics education and summarize instructors' accounts. The overview groups 122 distinct works into four categories, with further subcategories. A preliminary framework developed from existing review literature informed the interview guide. We interviewed 12 instructors from the University of Waterloo's Institute for Quantum Computing and the Perimeter Institute, who have collectively taught over 100 quantum courses. The accounts describe difficulties involving instructional representations, prerequisite knowledge, and the interpretation of students' reasoning, alongside reported diagnostic and instructional responses. We draw these accounts and the literature together as questions for instructional planning. The report does not resolve the intended classical--quantum comparison or establish comparative instructional effectiveness.
\end{abstract}

\maketitle

\section{Introduction}

Students often enter physics classrooms with well-established ideas about how the physical world works. Some of these ideas conflict with accepted scientific explanations and can persist even after formal instruction. Physics education research has commonly described such ideas as misconceptions, although they need not constitute stable or internally coherent mental models. Nor is all prior knowledge an obstacle to learning: preconceptions are simply ideas held before instruction, and students’ existing knowledge can provide productive resources for developing more sophisticated understanding~\cite{smith1994misconceptions}. Accordingly, a substantial body of research has sought both to identify common misconceptions and to develop instructional strategies that support their revision~\cite{gilbert1983concepts, scott1991teaching, soeharto2019review}.

An open question is whether classical and quantum physics courses demand distinct strategies for identifying and addressing misconceptions. Both domains require instructors to elicit students' reasoning and help them examine its limits, but quantum concepts can pose different representational and mathematical demands. This raises the question of how instructors describe the similarities and differences between addressing misconceptions in classical and quantum physics.

This report describes how experienced quantum physics instructors explain misconceptions and the strategies they use to identify and address them. The interview guide also explored comparisons with classical instruction. A preliminary framework developed from review literature informed semi-structured interviews with 12 instructors at the University of Waterloo's Institute for Quantum Computing and the Perimeter Institute, who collectively have taught over 100 courses. We situate their accounts within a teaching-oriented organization of 122 selected works.

\Needspace{6\baselineskip}
The questions underlying the inquiry are restated here to reflect the descriptive scope of this report: \begin{itemize}
\item [\textbf{RQ1:}] What sources of misconceptions, diagnostic methods, and remediation strategies are described in the selected literature, and how can they be organized for instructional use?
\item [\textbf{RQ2:}] What strategies do participating instructors report using to identify and address misconceptions in quantum physics, and how do they judge their usefulness?
\item [\textbf{RQ3:}] How do instructors compare the misconceptions encountered, and the approaches used to address them, in classical and quantum physics? \end{itemize}
This report addresses RQ1 and RQ2 descriptively. RQ3 motivated the inquiry and was raised in the interviews, but this revision does not present a dedicated synthesis of participants' classical--quantum comparisons. The earlier inference that instructors find similar strategies effective across the two domains is not retained: reported use and perceived usefulness of familiar approaches in quantum instruction do not, by themselves, answer the comparative question. RQ3 therefore remains unresolved in this report. This is a limitation of the report's scope, not a finding of equivalence or difference between the domains. Separately, the study did not measure comparative instructional effectiveness.

To situate our study, we first highlight key contributions in the literature on student misconceptions in quantum physics instruction. A foundational review by Singh and Marshman (2015)~\cite{singh2015review} synthesized research on common difficulties in upper-level quantum mechanics, emphasizing prevalent reasoning errors, conceptual overgeneralizations, and the importance of research-driven instructional strategies. Complementing this, a 2017 review by Krijtenburg-Lewerissa et al.~\cite{krijtenburg2017insights} examined teaching practices at the secondary and introductory undergraduate levels, identifying shared misconceptions, gaps in pedagogical approaches, and a lack of validated assessment tools.

Since these reviews, the field has seen a marked increase in targeted studies. Several have focused on specific conceptual challenges students face, such as difficulties with separation of variables~\cite{tu2020students}, solving bound and scattering state problems~\cite{tu2021students}, and applying delta functions in quantum mechanics~\cite{tu2023students}. Parallel work has explored strategies for conceptual remediation across a range of foundational topics, including single-photon interference in Mach-Zehnder interferometers~\cite{marshman2017investigating}, the formalism and postulates of quantum mechanics~\cite{marshman2019validation}, the interpretation of the double-slit experiment~\cite{sayer2017quantum}, and degenerate perturbation theory in the context of the Zeeman effect~\cite{keebaugh2019improving}. These efforts have contributed to the development of research-based tools such as inquiry-driven tutorials, interactive simulations, and conceptual surveys designed to enhance learning and measure instructional effectiveness. In addition, some studies have adopted epistemological frameworks to analyze how students frame and engage with quantum mechanics problems~\cite{modir2017students, modir2019framing}.

Related research has already examined instructors' perspectives. Siddiqui and Singh surveyed twelve faculty members about quantum-mechanics course goals, student preparation, teaching challenges, and approaches~\cite{siddiqui2017diverse}. Meyer et al. combined a survey of quantum-information-science instructors with follow-up interviews, identifying concerns about mathematical preparation, diverse backgrounds, and course design~\cite{meyer2022interdisciplinary}. A related analysis of the same interview project examined media hype and instructors' responses to students' preconceptions~\cite{meyer2023media}. More recently, Kruse, Griston, and Wilcox characterized teaching methods, resources, and content coverage using responses from 76 instructors~\cite{kruse2025characterization}.

The present study focuses on how instructors in one institutional setting explain misconceptions, diagnose them, and judge their instructional responses. It relates those accounts to a teaching-oriented organization of the literature without comparing measured instructional effectiveness.

The rest of this paper is organized as follows: Section~\ref{sec:method} explains the methodology for identifying relevant literature and conducting instructor interviews. In Sec.~\ref{sec:relationship}, we clarify the distinction between misconceptions and other sources of misunderstanding. Section~\ref{sec:review} details our classification scheme for categorizing the sources of misconceptions, diagnosis methods, and remediation strategies, with illustrative examples. Section~\ref{sec:Interviews} then presents the interview findings, highlighting instructors’ insights into the nature of quantum misconceptions and their judgments of the usefulness of various instructional strategies. Finally, Section~\ref{sec:Discussion} brings the accounts and selected literature together to discuss instructional implications and identify open questions.

\section{Method}\label{sec:method}

\subsection{Literature search}\label{sec:search}

We assembled a teaching-oriented overview of selected literature using database searches and citation tracking. The \textit{Web of Science} search was conducted between December 2023 and January 2024 using the following string:
\begin{quote}
ALL = ("misconception" OR "alternative conception" OR "preconception" OR "conceptual misunderstanding") AND ALL = ("physics")    
\end{quote}
The search returned $419$ results. The first exclusion criterion was
\begin{itemize}
    \item The article had not been published in a peer-reviewed journal.
\end{itemize}
This reduced the list to $219$ entries. The author then screened these records using the following exclusion criteria:
\begin{itemize}
    \item The article is not in the field of physics education research.
    \item The article is unrelated to addressing misconceptions.
\end{itemize}
These criteria reduced the list to $102$ entries. To identify overlooked studies, we employed forward and backward citation tracking:
\begin{itemize}
    \item Backward snowballing: Reviewing references of selected studies.
    \item Forward snowballing: Using Google Scholar’s citation function to find newer papers citing the selected studies.
\end{itemize}
Backward and forward citation tracking and related reading in March 2024 expanded the working tabulation from $102$ to $126$ entries. A subsequent audit identified a reference cited twice with different summaries. The incorrectly matched additional row was removed, leaving $125$ distinct works. A further scope and verification check during revision removed the Krist (1992) entry because the original article or abstract was unavailable for verification of its substantive summary. The language-learning study by Mizushina and Magara (2007) and the school-geometry study by Dembo, Levin, and Siegler (1997) were also removed as peripheral to the report's instructional focus. The present database therefore contains $122$ distinct works. A complete source-by-source record of the discovery route was not retained for every addition. The works are sorted into four categories: Sources of misconceptions, Diagnostic methods, Remediation strategies, and Other. The categories are further subdivided into subcategories (see Fig.~\ref{fig:organizational_scheme}). Each work, its subcategory, and a brief summary are provided in Appendix \ref{add:database}.

The author classified each work by reading its title and abstract and, where necessary, relevant sections of the full text. Each work was assigned a primary category according to its principal pedagogical contribution, while secondary themes were recorded in notes. No independent second reviewer performed the classification. Literature-summary and category checks during revision used the AI assistance described in the Acknowledgments; this did not involve analysis of the interview transcripts. 

Some non-journal, school-level, and adjacent-discipline sources were deliberately retained for their relevance to teaching practices or conceptual learning. Other departures from the initial criteria resulted from inconsistent screening. The broader collection is retained to illustrate approaches university instructors may consider, not to treat findings from other settings as direct evidence about university quantum instruction. It is therefore presented as an overview of selected literature rather than a comprehensive systematic review.

The instructor-perspective studies cited in the introduction~\cite{siddiqui2017diverse,meyer2022interdisciplinary,meyer2023media,kruse2025characterization} were added as context during manuscript revision, not as entries in this database. In particular, the 2025 survey postdates the original collection. These citations and the historical course-catalog sources do not change the 122-work total in Appendix~\ref{add:database}.

\begin{figure*}[t]
    \centering
    \includegraphics[width=\textwidth]{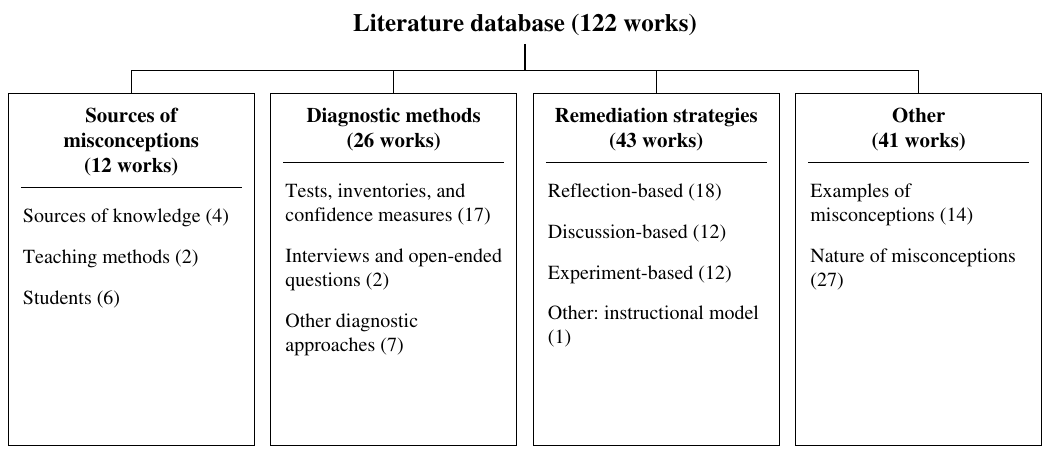}
    \caption{Overview of our organizational scheme for the literature on misconceptions in physics. Numbers indicate the count of distinct works in each category. For brief summaries of each work, see Tables~\ref{table:sources} to~\ref{table:nature} in Appendix~\ref{add:database}. The diagram was rendered with AI-assisted code from the tabulated counts.}
    \label{fig:organizational_scheme}
\end{figure*}

\subsection{Interviews}\label{sec:interviewmethod}

\subsubsection{Participant selection}

We focused on the University of Waterloo due to its recognized strengths in quantum physics education and its affiliation with the Perimeter Institute. The selected historical catalog in Appendix~\ref{app:courses} contains 13 undergraduate title groups on quantum physics, 33 graduate title groups, and eight undergraduate title groups covering quantum physics. Identical titles within each level are grouped for presentation; these counts are not counts of independently verified course offerings. Nineteen potential participants were identified through university course listings and departmental information, and three through referrals. All 22 were contacted by email using a standardized invitation; twelve participated and ten did not respond. Recruitment ended after this invitation and referral round; no formal saturation assessment was performed. Interviews took place in January--February 2024.

The 12 participants included nine faculty members at the time of the interviews and three experienced guest lecturers for summer schools. Each participant had taught quantum physics in a university setting for a minimum of five years, ensuring substantial teaching experience. The accounts concerned predominantly undergraduate teaching; three participants drew on graduate-course experience and three on summer-school experience. These counts do not imply mutually exclusive teaching-context groups.

At the time of the interviews, the author was a PhD student at the institution where all participants taught. Prior relationships ranged from no previous interaction to collaborative and supervisory relationships; the author was not evaluating participants' teaching. Shared disciplinary knowledge facilitated detailed technical discussion but could also leave assumptions unstated. Follow-up questions therefore sometimes asked instructors to explain why an answer was problematic, what they thought a student believed, or why a teaching response was useful.

\subsubsection{Interview protocol}

A preliminary framework developed in September 2023 from existing reviews of misconceptions, conceptual change, diagnostic approaches, and quantum-physics learning informed the interview guide~\cite{gilbert1983concepts, scott1991teaching, soeharto2019review, singh2015review, krijtenburg2017insights}. Its five broad areas corresponded to the core questions below; the detailed categories in the completed literature analysis were not all fixed at that stage.

We designed our interview protocol following guidelines for qualitative interviewing \cite{busetto2020use, dilley2000conducting, brod2009qualitative, solarino2021challenges}. The final protocol included five core, open-ended questions addressing:
\begin{itemize}
    \item What are university students' common misconceptions about quantum physics?
    \item What are the sources of these misconceptions?
    \item What are your strategies for identifying misconceptions?
    \item What are your strategies for addressing misconceptions?
    \item Besides misconceptions and gaps in knowledge, what else causes students to make mistakes?
\end{itemize}
All twelve participants were explicitly asked whether identifying and addressing misconceptions in quantum physics required different approaches from classical physics. Five had introduced a comparison spontaneously before that prompt; the remaining seven discussed the comparison in response to it. Follow-ups asked how instructors knew whether an approach had helped or why they regarded it as useful. Participants were not shown the literature framework.

We asked follow-up questions as relevant, while ensuring all core questions were covered to maintain consistency across interviews. Each interview was scheduled for 30 minutes via Zoom, though many lasted 45 minutes or longer based on participant engagement.

Interviews were audio-recorded and transcribed with TurboScribe. The author checked each transcript against its recording, paying particular attention to technical terminology, verbally described equations or symbols, and unclear passages. Errors affecting meaning were corrected, and passages that remained unclear were marked as uncertain rather than reconstructed. All participants provided recorded verbal consent. Identifying information was removed from the working transcripts to protect the confidentiality of statements. Participation was not wholly anonymous: eleven instructors are named in the acknowledgments, while a twelfth requested anonymity. 

\subsubsection{Interview analysis}\label{sec:interviewanalysis}

The initial analysis took place in February--March 2024. The author read all twelve checked transcripts in full and marked passages or exchanges relevant to the five interview areas, retaining the surrounding context. A spreadsheet linked participant identifiers and passages or transcript locations to brief descriptions and notes comparing similar or contrasting comments. More specific topics developed within the broad interview areas through comparison across participants.

Observations were included when they addressed a research question and their meaning was clear and supported by the transcript. Related passages were compared before being described as shared findings. Ambiguous statements were not used to support stronger interpretations unless the surrounding discussion clarified them. Differences and exceptions were retained to qualify findings or describe variation rather than being merged into a single position.

The author performed the substantive interpretation. The supervisor discussed the organization of the developing findings and challenged some interpretations but did not independently code the transcripts. These discussions prompted returns to the transcripts and, in some cases, narrower wording. The author also revisited the transcripts during drafting, separating some initially combined observations and identifying further qualifications. Findings remained linked to their supporting passages; explanatory material added by the author is distinguished from interview observations in Section~\ref{sec:Interviews}.

\subsubsection{Strengths and limitations of the interview approach}

Interviews were used to examine instructors' accounts of recurring difficulties and reported instructional responses, rather than to diagnose students directly. One strength is that lecturers, through years of teaching, develop an intuitive understanding of common student struggles, providing a broad overview of recurring challenges. Responsive follow-up questions allow interviewers to probe accounts of assessment errors, classroom engagement, and recurring requests for help in ways that fixed-response instruments do not. Moreover, interviews can be more time-efficient than large-scale student surveys or detailed diagnostic assessments. However, this method has notable weaknesses. Lecturer recollections of student difficulties may be influenced by selective memory or biases. The identification of challenges is subjective and may vary between lecturers based on their personal experiences, teaching styles, and interactions with students. Furthermore, some student difficulties may not be apparent to lecturers, particularly if students do not actively express them in class. Lastly, because this method does not directly involve students, it risks overlooking nuanced challenges that students themselves perceive but lecturers do not recognize.

Comparisons across transcripts, attention to exceptions, and returns to the original passages provided checks on interpretation. The literature supplied context for the interview accounts rather than independent verification of the student difficulties or outcomes described. The findings remain accounts from a particular institutional setting, not estimates of prevalence or comparative instructional effectiveness.

\section{Defining ``Misconception''}\label{sec:relationship}

Before presenting our main results, we clarify how we use the term ``misconception'' and distinguish it from other sources of student difficulty. Although early work highlighted substantial variation in how student misconceptions were conceptualized~\cite{gilbert1983concepts}, contemporary usage generally refers to student ideas that conflict with accepted scientific accounts, while differing theoretical perspectives remain on the nature and organization of those ideas~\cite{docktor2014synthesis}. In this study, we use ``misconception'' operationally to denote an idea or interpretation that is inconsistent with the accepted scientific account in the context under consideration, rather than an isolated arithmetic error or an incorrect answer alone. This usage does not imply that misconceptions form a uniform or sharply defined category~\cite{smith1994misconceptions}. To organize the discussion that follows, we review two established typologies and adopt a broad pedagogical distinction between misconceptions and gaps in prior knowledge. This distinction is intended to help characterize instructional responses rather than to provide an exhaustive taxonomy of student thinking, and the two categories may overlap. In reporting our interview results, moreover, we use ``misconception'' to refer to instructors' accounts of student thinking rather than to conceptions independently diagnosed in students.

One prominent typology was proposed by the National Research Council ~\cite{national1997science}. They presented five types of misconceptions:
\begin{itemize}
\item [(1a)] Preconceived notions: Intuitive but often incorrect understandings of scientific concepts based on everyday experiences.
\item [(1b)] Non-scientific beliefs: Views learned outside the scientific community that often conflict with scientific evidence.
\item [(1c)] Conceptual misunderstandings: Faulty explanations formed when instruction leaves conflicts with prior ideas unresolved.
\item [(1d)] Factual misconceptions: Incorrect facts often learned early and retained without being challenged.
\item [(1e)] Vernacular misconceptions: Misunderstandings due to the different meanings of words in scientific versus everyday language.
\end{itemize}
Liu and Fang's literature review~\cite{liu2016student} presents another classification. Their work reports reviewing 135 studies on misunderstandings about force and acceleration across various educational levels, identifying 38 misunderstandings related to force and 15 related to acceleration.\footnote{The reference list in Liu and Fang ends at Ref.~60, but the authors report 135 reviewed studies and explain that they cite representative papers rather than every relevant work found.} Reported explanations were grouped into four categories:
\begin{enumerate}
\item [(2a)] {Incomplete or partial understanding:} Occurs when students do not fully comprehend key concepts, leading to gaps in their knowledge.
\item [(2b)] {Preconceived misunderstandings:} Misunderstandings carried over from a student’s life experiences.
\item [(2c)] {Wrong interpretations and comprehensions:} When students understand a concept in isolation but are unsure how to apply it.
\item [(2d)] {Vernacular misunderstandings:} Arise from students interpreting scientific terminology based on everyday language usage.
\end{enumerate}
Our category of ``gaps in knowledge'' broadly corresponds to 2a, whereas categories 1a--1d and 2b--2c may involve misconceptions. Categories 1e and 2d are less clear-cut: unfamiliarity with a technical term may indicate a gap in knowledge, while associating that term with an incorrect explanatory idea may indicate a misconception. The two can also coexist, so a linguistic error alone is insufficient to determine which is present or how readily it can be addressed.

\section{Categorizing the literature}\label{sec:review}

This section addresses \textbf{RQ1} by organizing the literature into broad categories intended to help instructors navigate common themes. The categories provide a teaching-oriented guide, rather than evidence of instructional effectiveness. We discuss sources of misconceptions in Section~\ref{sub:source}, diagnostic methods in Section~\ref{sub:distinguish}, and remediation strategies in Section~\ref{sub:strategies}, using selected examples to illustrate each area. Works concerned with examples and the nature of misconceptions are grouped under ``Other''; this category does not represent an additional instructional strategy.

\subsection{Sources of misconceptions}\label{sub:source}

The literature on sources of misconceptions is summarized in Table~\ref{table:sources}. For organizational purposes, we group the identified sources into three broad, non-exclusive categories:
\begin{itemize}
\item {Sources of knowledge:} Misconceptions can arise from inaccurate information from sources students believe they should trust to some extent. This includes teachers themselves, textbooks, videos, and other literature.
\item {Teaching methods:} Ineffective teaching strategies can contribute to forming and perpetuating misconceptions. This includes complicated entry points into a subject or poorly designed teaching activities. 
\item {Students:} Even in the presence of accurate sources of knowledge with sound teaching methods, students can develop misconceptions. This can be from their prior experiences, incorrect reasoning, stage of development, abilities, or interest in the subject.
\end{itemize}
Our three-bin framework broadly aligns with that of Ref.~\cite{suprapto2020we} when ``Textbooks and Literature'' are grouped with ``Teachers.'' For pedagogical clarity, we classify teachers, textbooks, and other media as sources of knowledge; teaching methods as instructional factors; and student-related characteristics as learner factors. These categories are not mutually exclusive. An ambiguous diagram, for example, may reflect both the content presented and the way it is presented. We therefore assign each work to a single primary category for organizational purposes, without implying that only one source contributes to the difficulty. These placements indicate possible contributors or relevant conceptual-change accounts, not established causal effects. Works primarily cataloguing misconceptions are listed separately among the examples in Table~\ref{table:examples}.

When students do not yet understand a subject, they may have difficulty distinguishing correct from faulty sources of knowledge. Kulgemeyer and Wittwer examined the emergence of misconceptions during self-directed learning, particularly through the use of physics explainer videos, which have become increasingly popular over the past two decades~\cite{kulgemeyer2023misconceptions}. Their study assessed the influence of these videos on students' understanding by dividing 149 physics learners into two groups. One group viewed a video containing misconceptions about the concept of force, while the control group watched a scientifically accurate video. Both videos were similar in terms of comprehensibility and duration, differing primarily in content accuracy. The results revealed that the group exposed to misconceptions developed a comparable level of perceived understanding to the control group but acquired more misconceptions and less scientific knowledge. This finding highlights the critical importance of evaluating the accuracy of educational content in textbooks and other learning resources to prevent the dissemination of misinformation.

Teaching approaches that help fill knowledge gaps can also contribute to misconceptions if they are poorly implemented. Cook's discussion of visual representations illustrates this concern~\cite{cook2006visual}. Visual aids can create difficulties when students lack the prior knowledge needed to interpret them or when the representations overwhelm their cognitive capacity. In such cases, students may focus on superficial features and draw incorrect conclusions. Careful visual design and the use of multiple representations may help mitigate these difficulties. Our interviewees described concerns about representations in quantum physics that are consistent with this account, although their observations do not establish cognitive overload or abstraction as the cause.

\subsection{Diagnostic methods}\label{sub:distinguish}

Attention to possible sources may help prevent misconceptions, but instructors still need ways to investigate students' reasoning. Table~\ref{table:diagnostic} summarizes the diagnostic approaches identified in the literature, including interviews, open-ended questions, concept inventories, and tests with confidence ratings or multiple tiers. For ease of use, we group tests and inventories together, distinguish interviews and open-ended questions, and place analytic and computational methods under other diagnostic approaches. Each work appears under the heading that best represents the contribution emphasized here. These groups are intended to guide instructional choices, not to separate mutually exclusive methods: interviews may be used to develop tests, and the same analysis may be applied to several instrument formats. Interviews generally require more individual contact, whereas fixed-response instruments can be easier to administer at scale, but neither is restricted to a particular class size.

\subsubsection{Interviews and open-ended questions}

Individual interviews allow students to explain their reasoning and give the interviewer opportunities for follow-up questions. They can reveal interpretations that are not apparent from an answer alone, but conducting and analyzing them is time-intensive. The interviewer's questioning and interpretation also affect the account obtained, while trust can support candid discussion.

Open-ended written questions likewise let students express their ideas without predefined answer choices. As with interviews, interpreting the responses takes time and judgment, and their usefulness depends on students providing enough detail to reveal their reasoning.

Concept maps offer another way to elicit relationships between ideas. Blizak and Chafiqi used them to diagnose university students' misconceptions about light~\cite{djanette2014determination}.

\subsubsection{CRI and multi-tier tests}

Hasan et al.~\cite{hasan1999misconceptions} applied the Certainty of Response Index (CRI) in undergraduate classical mechanics, asking students to rate their confidence in multiple-choice answers from 0 (total guess) to 5 (complete confidence). They used these ratings to help interpret the class's understanding. An incorrect answer given with low confidence may indicate a knowledge gap, whereas one given confidently may suggest a misconception; neither diagnosis can be established from confidence alone~\cite{WOS:000836287700001}. This answer-plus-confidence format differs from the answer-plus-reason format commonly known as a two-tier diagnostic test, examples of which predate Hasan et al.'s study~\cite{gurel2015review}.

Confidence ratings do not reveal why an answer was chosen: students may confidently select a correct answer for an incorrect reason. Multi-tier tests address this limitation by also eliciting reasoning~\cite{gurel2015review}. A common two-tier format asks for an answer and a reason, while a three-tier format adds a confidence rating. Four-tier tests ask students to rate their confidence in the answer and the reason separately~\cite{yuberti2020four, diani2019four}, although the order of these components varies across instruments. The additional responses can help distinguish the reasoning behind identical answers, but do not guarantee a valid diagnosis. Conversely, students may understand a concept yet lack confidence in their answer. Recognizing this group separately may help instructors identify the support they need~\cite{fadllan2019analysis, saglam2015confidence}.

Three reviews illustrate the range of diagnostic formats used in the literature~\cite{gurel2015review, soeharto2019review, resbiantoro2022review}. Gurel et al.~\cite{gurel2015review} examined 273 publications from 1980 to 2014 and found interviews in 53\%, open-ended tests in 34\%, ordinary multiple-choice tests in 32\%, and multi-tier tests in 13\%. The percentages overlap because a study could use more than one instrument. In a review of 111 studies from 2015 to 2019, Soeharto et al.~\cite{soeharto2019review} reported approximately 11\%, 24\%, 32\%, and 33\%, respectively, for the same four formats. Resbiantoro et al.~\cite{resbiantoro2022review} likewise identified several formats in 72 articles published between 2005 and 2020. These reviews differ in scope and categorization, and their coverage periods overlap. Their percentages therefore cannot, on their own, establish a field-wide shift in diagnostic methods or differences in effectiveness.

\subsection{Remediation strategies}\label{sub:strategies}

Identifying and addressing misconceptions can be an iterative process. We use remediation to include refining and reorganizing existing ideas as well as replacing them, drawing on conceptual-change research~\cite{scott1991teaching,smith1994misconceptions}. For pedagogical clarity, we organize the strategies into three broad groups: reflection-based, discussion-based, and experiment-based approaches. These headings describe instructional emphases rather than mutually exclusive mechanisms; predict--observe--explain activities, for example, can combine all three. Each work is listed once under a primary heading, with an additional instructional-model category for work that does not fit readily into these groups. Table~\ref{table:strategies} gives examples of the three main approaches, and Table~\ref{table:remediation} lists all works in the remediation category.

\begin{table*}[ht]
\centering
\caption{Summary of educational strategies categorized into reflection-based, discussion-based, and experiment-based approaches.}
\label{table:strategies}
\renewcommand{\arraystretch}{1.12}
\begin{tabularx}{\textwidth}{lX}
\toprule
\textbf{Strategy} & \textbf{Explanation} \\ 
\midrule
\textbf{Reflection-based} & \\ 
Concept Mapping & Learners create maps linking concepts by connecting words or phrases, illustrating relationships between ideas. This is intended to help learners organize and structure their knowledge. \\ 
Journaling & Learners record their thoughts, understanding, and progress over time to support reflection. \\ 
Predict-Observe-Explain & Learners make predictions about an outcome, observe an experiment to see what actually happens, and then explain the results. This invites learners to reconsider their understanding in light of observed evidence. \\
Conceptual-Change Text & Learners are presented with text that challenges existing beliefs and encourages cognitive conflict. These texts are designed to support revision toward scientifically sound concepts.\\ 
\midrule 
\textbf{Discussion-based} & \\ 
Group Discussions & Learners participate in structured or informal discussions among learners or instructors to explore ideas, clarify understandings, and negotiate interpretations. These discussions are intended to encourage critical examination of ideas through dialogue. \\ 
Listening Activities & Learners listen to others discuss concepts. These can be pre-recorded conversations or live ones in class. Contrasting ideas give learners an opportunity to reconsider their own reasoning. \\ 
\midrule
\textbf{Experiment-based} & \\ 
Laboratory & Learners engage in hands-on experiments in a laboratory setting, enabling them to engage with processes relevant to the subject matter directly. This gives learners opportunities to connect theory with practical observations. \\ 
Simulations & Learners use computer simulations to mimic real-world processes or experiments, allowing them to interact with complex systems in a controlled, virtual environment. This allows learners to explore models and test predictions without constructing the corresponding physical apparatus. \\ 
\bottomrule
\end{tabularx}
\par\smallskip\noindent\begin{minipage}{\textwidth}\small\RaggedRight The descriptions summarize intended instructional functions; outcomes depend on activity design and implementation.\end{minipage}
\end{table*}

\subsubsection{Reflection-based strategies}

Reflection-based strategies encourage students to introspect and critically reassess their knowledge, integrating new and accurate information. These methods, often involving individual activities, can also include collaborative elements. Examples of reflection-based strategies include concept mapping~\cite{WOS:000318278800002}, journaling~\cite{hein1999using}, model-based teaching~\cite{ogan2007effects}, predict-observe-explain activities~\cite{miller2013role}, and conceptual change text~\cite{baser2007effectiveness, csahin2010computer, durmucs2010effects, franco2012examining}.

Hein's 1999 article described a ``folder activity'' for identifying and addressing misconceptions through writing~\cite{hein1999using}. The activity was introduced in an algebra-based introductory physics course for non-science majors at American University in Washington, DC, which often enrolls students with limited mathematical skills and no prior physics education. Used 5--10 times per semester, it presents short-answer prompts about course concepts that students address using their class notes. Some prompts also ask students to create multiple-choice questions and justify their answers. Hein argues that this process helps students clarify their understanding while allowing instructors to identify common misconceptions for further discussion, making it useful for both diagnosis and remediation.

Reflection activities can also involve brief predictions rather than extended writing. Miller et al.~\cite{miller2013role} studied the predict-observe-explain approach through 22 mechanics and electricity-and-magnetism demonstrations in introductory courses at two universities. They recorded students' predictions and collected reports of the demonstrations within one day and at the end of the semester. Among students who made a prediction, approximately 18\% of reports collected within one day did not match the observed outcome, although the mismatch alone does not identify its cause. Students who made predictions reported the outcomes more accurately, whether or not their predictions were correct. Prediction-making was also associated with better retention, but because students chose whether to make predictions, the study cannot isolate its causal effect.

\subsubsection{Discussion-based strategies}

Discussion-based activities invite students to encounter different perspectives and reasoning through interaction with peers and instructors. Positive results have been reported for in-person discussions~\cite{leinonen2013overcoming} and for listening to recorded conversations~\cite{muller2007tackling}. These benefits are not universal, however: Wendt and Rockinson-Szapkiw found increased misconceptions in their online-collaboration condition~\cite{wendt2014effect}.

Classroom discussion can range from informal exchanges to structured activities. Leinonen et al.~\cite{leinonen2013overcoming} examined a one-hour session in an introductory thermal physics course at the University of Eastern Finland. Students worked individually, received hints, and then discussed their reasoning with peers. The researchers used written explanations and peer-discussion recordings to examine understanding; the reported analysis focused on 65 students who participated in both the intervention and the course examination. For example, scientifically correct explanations for work in an isobaric process increased from 52\% during individual work to 80\% after peer discussion. Because the activities formed a sequence, the results do not isolate the contribution of peer discussion alone.

\subsubsection{Experiment-based strategies}

The final category, experiment-based activities, emphasizes ``hands-on'' engagement, allowing students to interact directly with physical or simulated systems. These experiences provide opportunities to visualize concepts, receive feedback, and reconsider ideas in light of observed outcomes. Experiment-based learning includes physical models and demonstrations, such as water-circuit analogies~\cite{korganci2015importance}, as well as simulation-based activities. These simulations include interactive tools~\cite{myneni2013interactive, dutt2012decisions, falloon2019using, hockicko2014correcting, phanphech2019explaining, schneps2014conceptualizing} and virtual reality technology~\cite{kozhevnikov2013learning}.

Simulations can incorporate instructional support as well as opportunities for exploration. The Virtual Physics System (ViPS), developed by Myneni et al.~\cite{myneni2013interactive}, combines pulley simulations with tutoring designed to address common misconceptions. Its process has three stages:
\begin{enumerate}
    \item Pre-Test Phase: Students answer problems about pulleys, providing a baseline for identifying possible misconceptions.
    \item Tutoring Phase: ViPS uses this information to guide problem-solving, offering real-time feedback and hints and adjusting support to the student's performance in accordance with the zone of proximal development principle~\cite{shabani2010vygotsky}.
    \item Post-Test Phase: Students answer questions after instruction so that changes in their responses can be assessed.
\end{enumerate}
A dynamic student model tracks progress throughout the activity, allowing feedback and challenges to adapt to each learner. This individualized support extends what is typically possible in laboratory-based experiments.

Myneni et al.~\cite{myneni2013interactive} evaluated ViPS in a virtual-only condition and in a separate comparison of virtual--physical and physical--virtual sequences. The virtual-only group improved from pretest to posttest. In the sequence comparison, students beginning with ViPS had a larger initial gain than those beginning with physical pulleys, but the final gains did not differ significantly between sequences. These results concern the combined simulation-and-tutoring activity, not the simulation medium alone.

\section{Interviews with quantum physics educators}\label{sec:Interviews}

Many tools for identifying and addressing misconceptions were developed in classical physics contexts. To explore how instructors view their use in quantum physics, we conducted long-form interviews with quantum physics instructors at the University of Waterloo. The following sections organize instructors' accounts by the main topics covered in the interviews. Most summaries combine comments from several instructors and should not be read as unanimous positions or as the reasoning of a single participant. They describe perceived difficulties and proposed explanations, rather than measured prevalence or established causes. Where needed, we add physics clarifications to distinguish the reported difficulty from the relevant physical account; these clarifications are not additional interview observations. These author-supplied explanations do not imply that participants reviewed or endorsed the revised scientific account.

\subsection{Conceptual difficulties reported by instructors} 

\textit{Entanglement.---}Instructors reported that some students interpret entanglement as a means of faster-than-light communication, believing that a controllable local action on system $A$ could instantly send a message to its entangled partner. They associated this idea in part with portrayals in popular educational materials. Although measurements can support valid conditional predictions about the partner's spin, local trace-preserving operations leave its unconditional measurement statistics unchanged. The correlations therefore do not permit faster-than-light signaling.

\textit{Spin.---}Instructors described students interpreting spin as literal physical rotation, for example by imagining an electron as a spinning top. They linked this difficulty to educational depictions of electrons as rotating charged spheres. Spin is an intrinsic property of particles, not the rotation of a classical charged sphere.

\textit{Observer effect.---}Instructors reported the belief that wave-function collapse requires a conscious observer, associating it with textbook illustrations of eyeballs and portrayals in popular media. Standard measurement models describe interactions with measurement devices without requiring consciousness; how to interpret state reduction is a separate question.

\textit{Uncertainty in measurement outcomes.---}Instructors described confusion between uncertainty about a system's state and the spread of measurement outcomes for an observable in a specified state. Even a precisely specified pure state can yield uncertain outcomes, whereas an energy eigenstate has zero energy variance. The preparation uncertainty relation concerns the spreads of noncommuting observables in a given state; it neither makes every observable uncertain nor imposes a general limit on the accuracy of every measurement.

\textit{Density matrices.---}Instructors described difficulties distinguishing uncertainty about how a system was prepared from the probabilities of quantum measurement outcomes. To clarify the physics, we note that density matrices describe both pure and mixed states. A pure state need not yield a definite value for every observable, while a mixed state may result from a randomized preparation or arise as the reduced state of an entangled system. Thus, a pure state does not imply certainty, nor can a mixed state be equated simply with classical ignorance.

\textit{Discreteness and quantization.---}Instructors described confusion about whether discreteness is unique to quantum mechanics. To clarify the physics, we distinguish discrete allowed frequencies from quantized energy levels. Bounded classical wave systems can have discrete normal-mode frequencies, but the energy of an ordinary classical harmonic oscillator, $E=\tfrac12 m\omega^2 A^2$, varies continuously with amplitude.

\textit{Wave-particle duality.---}Instructors associated some students' difficulties with textbook language suggesting that a quantum object is simultaneously a classical particle and a classical wave. Quantum theory must account for both interference patterns and localized detection events; neither feature makes one classical picture universally applicable.

\textit{Quantum theory of light.---}Instructors described students confusing the energy associated with a mode's frequency with the number of photons occupying that mode. Similar ladder-like diagrams for the two concepts may contribute to the confusion. For a mode of frequency $\nu$, $n$ photons contribute excitation energy $nh\nu$, so changing photon number is distinct from changing mode frequency.

\subsection{Proposed sources of conceptual difficulty}

The instructors collectively identified eight sources of misconceptions. The first four pertain to teaching methods, the next three to educational materials, and the last to the students.

\textit{Traditional introductions to quantum mechanics.---}Several instructors discussed the mathematical demands of beginning quantum mechanics with wave functions. Some favored introducing discrete systems and Dirac notation first as a way to reduce the initial mathematical burden.

Other comments concerned the conceptual demands of beginning with the Stern--Gerlach experiment, which introduces ideas such as spin and measurement collapse together. Favoring a discrete-state introduction does not necessarily imply support for starting with this experiment: notation, the choice of state space, and the introductory experiment are separate teaching decisions. The comments may also reflect different instructors' preferences rather than a single position.

\textit{Distinguishing experimental observations from interpretations.---}Instructors described confusion that can arise when teaching blurs the distinction between experimental predictions and interpretations of quantum mechanics. Students naturally ask what the theory says about physical reality, including why an ideal projective measurement updates a state to the eigenspace associated with its outcome. This description applies to ideal projective measurements, not to all measurements, and quantum theory generally predicts probabilities rather than certain individual outcomes. Discussions of interpretation can be valuable, but should distinguish empirical predictions from claims about their meaning. Otherwise, students may mistake an interpretative claim, such as the need for a conscious observer, for an established requirement of the theory.

\textit{Prolonged disconnect from physical systems.---}Although quantum mechanics requires substantial mathematical work, instructors expressed concern that a prolonged emphasis on formal manipulation can obscure its connection to physical systems. Students may then overlook the physical constraints on their calculations. We clarify the physics of this example as follows: in a superconducting ring, flux quantization depends on appropriate boundary and current conditions; it does not follow from Stokes's theorem for an arbitrary ring. Flux is quantized when a path with negligible supercurrent can be chosen inside the superconductor, whereas the more general quantized quantity is the fluxoid. The accounts also highlighted difficulties judging when quantum effects matter. Such judgments require more than a characteristic scale: whether a perturbation lifts a degeneracy depends on its matrix elements and symmetries, while interference depends on coherence and the measurement arrangement, not ion separation alone.

\textit{Jargon.---}One instructor who teaches students from several departments described technical language as a barrier for students from fields such as engineering and mathematics. As discussed in Section~\ref{sec:relationship}, this difficulty may involve unfamiliar vocabulary, an incorrect interpretation, or both. Clear explanations of technical terms may help, but difficulty with terminology alone does not establish a misconception.

\textit{Popular science media.---}Instructors described popular science media and other educational materials as possible sources of dramatized or oversimplified accounts of quantum phenomena. They most often attributed the belief that entanglement enables faster-than-light communication to popular media, suggesting that classroom discussions could help students examine the limits of those accounts.

\textit{Ambiguously worded textbooks.---}Instructors also identified ambiguous textbook language as a possible source of confusion. Describing a photon as both a wave and a particle, for example, can obscure the limits of those classical pictures. More precise wording and explicit discussion of those limits are possible instructional responses.

\textit{Ambiguous figures.---}Various ambiguous figures were mentioned by instructors. These include images of something rotating to depict spin, an eyeball beside a system to depict projective measurements, and rope-like drawings connecting systems to depict entanglement. 

\textit{Inappropriate use of problem-solving tools.---}One instructor described students applying a familiar problem-solving method in more complex settings without recognizing its limits. Inadequate prior instruction was also acknowledged as a possible contributor. Explicit discussion of a method's domain of validity is therefore a possible response.

\subsection{Diagnostic practices}

\textit{Identifying trends in formative assessment errors.---}Instructors described analyzing student responses to exercises to detect misconceptions. They used particular errors or patterns in student answers to identify areas of recurring difficulty.

\textit{Informal discussions with subsets of students.---}Instructors also described using informal conversations during class or office hours to investigate misconceptions. Although these conversations involved only a subset of students, instructors regarded them as useful for identifying issues to address with the entire class. They described spontaneous exchanges as revealing reasoning and misunderstandings that might not surface in formal assessments.

\textit{Compiling common misconceptions.---}One instructor described maintaining a list of misconceptions drawn from personal teaching experience and colleagues' observations. The instructor reported that similar difficulties recurred from year to year. A record of these patterns can guide preparation, but students' reasoning still needs to be checked to determine which ideas are present in the current class.

\subsection{Instructional responses and judgments of usefulness}

\textit{In-class quizzes with ``misconception traps''.---}The following synthesis brings together related polling activities described across the interviews rather than presenting a single procedure reported by every participant. Several instructors described quizzes containing ``trap answers'' based on common misunderstandings. Electronic clickers or polling apps reveal the distribution of responses, while attention to students' reasoning helps identify the ideas behind them. The activity can thus serve both diagnostic and instructional purposes. One instructor, for example, presents measurement scenarios to examine whether students believe a conscious observer is required for wave-function collapse.

Instructors described displaying poll results to show students that others had selected similar answers. In one reported approach, students are then asked to justify an answer they did not choose, encouraging them to consider alternative reasoning without having to defend their own response. These are the intended benefits of the activity, rather than outcomes independently assessed in this study.

The accounts also included peer discussion followed by another poll. Instructors regarded increased agreement on the correct answer as a useful immediate classroom signal. The interviews did not measure the frequency of successful remediation or assess independent reasoning, retention, or transfer.

\textit{Prediction-based class discussions.---}Another reported strategy asks students to predict the outcome of a scenario likely to elicit misconceptions and then discuss their reasoning. One instructor used an interferometry experiment, asking what would change if one path were blocked or its phase altered. The dialogue was intended to reveal students' predictions and prompt them to reconsider their ideas. One instructor noted, "I get them to the point where they don’t know what to believe anymore. Then, when they’re at that point, they’re ready to let go of their misconception." The account emphasizes eliciting an idea before challenging it.

\textit{Video demonstrations or live experiments.---}Three instructors described video demonstrations or, where possible, live experiments as useful for addressing misconceptions. These can be incorporated into predict-observe-explain activities or accompanied by direct explanations. A laser-and-slits demonstration, for example, illustrates classical wave interference rather than a uniquely quantum effect. Demonstrations may also engage students. Most instructors nevertheless reported limited options for demonstrations in quantum physics classrooms.

\subsection{Prerequisite knowledge gaps}

When asked about causes of mistakes other than misconceptions and knowledge gaps, three instructors pointed to deficiencies in prior knowledge. Their responses therefore returned to the knowledge gaps discussed in Section~\ref{sec:relationship}, rather than identifying a third category of error.

\textit{Insufficient mathematical background.---}Instructors reported gaps in second-year students' mathematical preparation for quantum mechanics. The subject demands an understanding of complex numbers, linear algebra, and a transition to abstract Hilbert spaces. They described these gaps as limiting students' ability to check their answers and apply quantum concepts.

\textit{Insufficient quantum physics background.---}Instructors described courses in which students encountered substantial quantum-mechanics content without sufficient background knowledge. Their examples included courses that do not exclusively teach quantum physics, such as statistical mechanics. One Waterloo course was described as almost evenly divided between classical and quantum statistical mechanics. The accounts linked these difficulties to the timing of preparatory coursework, rather than establishing a general problem with course scheduling.

\subsection{Instructional simplification and audience}

Several insightful side conversations emerged from the interviews, providing valuable perspectives on addressing misconceptions.

\textit{Instructional models and misconceptions.---}Instructors distinguished unintended difficulties from deliberately provisional simplifications. In the terms of Section~\ref{sec:relationship}, the latter are not necessarily misconceptions: using a model within its stated domain differs from believing it is universally valid. The progression from billiard-ball atomic models through the ``plum pudding'' model to orbitals and electron clouds also includes superseded historical explanations, not just approximations valid within a stated domain. Their instructional value may lie in examining predictions, evidence, and limitations rather than presenting each model as universally correct.

\textit{Prior ideas as resources.---}Some instructors argued that an initial, imperfect conception could provide a starting point for explaining why a different account is needed. They contrasted this with teaching students unfamiliar with the particular quantum idea. Unfamiliarity with that idea does not imply an absence of prior knowledge, and these were pedagogical judgments rather than learning outcomes established by the interviews.

\textit{Specialization in quantum physics.---}Instructors emphasized tailoring quantum mechanics education to the audience and its goals. Mathematics students, for example, may approach the subject through linear algebra and its additional physical rules. The accounts suggested that students could contribute to quantum algorithms without studying every application, such as scattering. This raises a curricular question about how to balance a common foundation with specialization, rather than establishing that a particular degree of specialization is necessary at the undergraduate level.

\begin{table*}[t]
\caption{Questions for instructional planning drawn from the selected literature and instructor accounts. The questions and possible responses are an author synthesis, not a validated intervention.}
\label{table:instructional_synthesis}
\centering
\begin{tabular}{@{}>{\raggedright\arraybackslash}p{0.23\textwidth}>{\raggedright\arraybackslash}p{0.40\textwidth}>{\raggedright\arraybackslash}p{0.29\textwidth}@{}}
\toprule
\textbf{Instructional question} & \textbf{Connection in this report} & \textbf{Possible response to consider} \\
\midrule
Is a representation being interpreted too literally? & Visual-representation literature and accounts of pictures used to depict spin and measurement (Sections IV A and V A--B)~\cite{cook2006visual}. & Ask students what the representation means and make the limits of the analogy explicit. \\[3pt]
Does an answer reveal the reasoning behind it? & Diagnostic literature and accounts of assessment errors, informal conversations, and polling (Sections IV B and V C--D)~\cite{WOS:000836287700001,gurel2015review}. & Examine reasons alongside answers before deciding how to respond. \\[3pt]
Does the difficulty involve terminology, prior knowledge, or an incorrect interpretation? & The distinction developed in Section III and accounts of jargon and prerequisite gaps (Sections V B and V E). & Clarify whether vocabulary support, prerequisite instruction, or discussion of an explanatory idea is needed; more than one may apply. \\[3pt]
Is a useful model being applied beyond its limits? & Conceptual-change discussion and accounts of method transfer and instructional simplification (Sections III, V B, and V F)~\cite{smith1994misconceptions,scott1991teaching}. & Make the model's assumptions and domain of validity explicit, and examine where its predictions cease to apply. \\[3pt]
\bottomrule
\end{tabular}
\end{table*}

\section{Discussion}\label{sec:Discussion}

Instructors reported finding familiar teaching strategies useful in quantum physics. The intended comparison in RQ3 remains unresolved; the reported use of familiar strategies does not establish how misconceptions or instructional approaches compare across the two domains. Instructors based their judgments on classroom experience, student explanations, assessment responses, changes during discussion, and repeated experience teaching the material. We bring these accounts together with the literature below to consider instructional implications and identify questions for further research, without treating instructors' judgments as evidence of comparative effectiveness.

\subsection{Integration of insights from literature and interviews}

The selected literature and instructor accounts can be brought together through four questions for instructional planning (Table~\ref{table:instructional_synthesis}). These questions are the author's pedagogical synthesis of material already discussed, not additional interview findings or validated decision rules. They overlap and are intended to guide consideration of a difficulty rather than classify it exhaustively.

Implementation also depends on resources. Individual discussions can provide depth, whereas fixed-response instruments may be easier to administer at scale. Interviewees' reports of limited options for quantum demonstrations identify a resource question, not a comparison of their use or effectiveness across domains.

The reported concerns about mathematical preparation and introductory choices overlap those discussed by Siddiqui and Singh~\cite{siddiqui2017diverse} and Meyer et al.~\cite{meyer2022interdisciplinary}; the comments about popular media also connect with the latter project's analysis of quantum-computing hype~\cite{meyer2023media}. Here, these concerns are organized around diagnosing and addressing misconceptions in the participating instructors' courses. The later survey by Kruse et al.~\cite{kruse2025characterization} documents variation across a larger set of courses, providing context rather than validation of these particular accounts.

\subsection{Directions for further research}

Multi-tier diagnostic research already covers quantum topics such as the photoelectric effect~\cite{WOS:000373840100003}. Building on this work, further studies could develop and evaluate question pools for additional topics and student populations, examining both instrument validity and instructional usefulness.

The difficulties described by educators at Waterloo are largely conceptual, consistent with Styer's emphasis on conceptual rather than operational difficulties~\cite{styer1996common}. Further research could examine how conceptual reasoning and mathematical preparation interact and which instructional responses address them, since these interviews do not establish their relative importance.

Instructors described gaps in mathematical training and foundational quantum physics as barriers in their courses, suggesting possible mismatches between prerequisites and course demands. Although these accounts do not establish a field-wide problem, they motivate closer examination of how mathematics and physics curricula can be aligned. Work on cohesive quantum-education programs, such as that of Asfaw et al.~\cite{asfaw2022building}, provides a useful starting point. Adaptable curricula could combine a common foundation with opportunities for specialization, reflecting the different goals of students preparing for careers in quantum physics and related fields.

Benefits reported for simulations in some classical physics settings motivate evaluation of their use for specific quantum-learning goals. Simulations of phenomena such as double-slit interference or entanglement may help students develop intuition, but their usefulness for those goals needs to be assessed. Tools such as IBM's Qiskit and Xanadu's PennyLane, which can simulate quantum algorithms, may likewise offer resources for quantum computing education.

The report brings together 122 selected works and instructors' accounts to help frame instructional decisions. A practical implication is to examine what an answer reveals about a student's reasoning before choosing a response. An unfamiliar term, missing prerequisite knowledge, an incorrect interpretation, and the overextension of a useful model may call for different forms of support and may coexist. Making these distinctions explicit connects the review's categories with the instructional concerns described in the interviews. These are considerations for planning instruction, not validated decision rules. The report does not establish the prevalence or causes of student misconceptions, determine comparative instructional effectiveness, or resolve the intended classical--quantum comparison.

\begin{acknowledgments}
I thank Bindiya Arora, Raffi Budakian, Richard Cleve, John Donohue, Alan Jamison, Brenda Lee, Kazi Rajibul Islam, Raymond Laflamme, Eduardo Martin-Martinez, Michael Reimer, and Christopher Wilson for participating in the interviews, along with a twelfth instructor who requested anonymity. I also thank Svitlana Taraban for her mentorship and valuable feedback on the manuscript.

This work was distinct from my PhD research~\cite{majidy2024noncommuting, majidy2023noncommuting, majidy2023critical, majidy2023unification, majidy2023non, yunger2022build, majidy2021detecting, majidy2019exploration, majidy2019violation, majidy2024building}. I acknowledge the general doctoral support provided by the Vanier CGS during the period in which this article was written.

After completing an earlier version of this report, the author revisited it in September 2026 with assistance from ChatGPT (OpenAI) to organize the text and clarify gaps in the reasoning. This assistance also included literature checks and summaries, review of category placements, scientific explanations, and LaTeX and figure code. It included editing manuscript passages reporting and discussing instructor accounts. The author directed the scope and approved the revisions, and remains responsible for the report. No interview recordings, complete transcripts, or participant-identity keys were supplied to ChatGPT. The reporting and synthesis of the instructor accounts, literature discussion, scientific explanations, and scope of the conclusions were revised. ChatGPT was not used to analyze the interview transcripts or generate participant-level findings.
\end{acknowledgments}

\par
\makeatletter
\let\bibliography@original@linkurl\hyper@linkurl
\def\hyper@linkurl#1#2{#1}
\edef\savedBibliographyInterlinePenalty{\the\interlinepenalty}
\interlinepenalty=10000 
\bibliography{apssamp}

@article{tu2020students,
  title={Students’ difficulties with partial differential equations in quantum mechanics},
  author={Tu, Tao and Li, Chuan-Feng and Zhou, Zong-Quan and Guo, Guang-Can},
  journal={Physical Review Physics Education Research},
  volume={16},
  number={2},
  pages={020163},
  year={2020},
  publisher={APS},
  doi={10.1103/PhysRevPhysEducRes.16.020163}
}

@article{tu2021students,
  title={Students’ difficulties with solving bound and scattering state problems in quantum mechanics},
  author={Tu, Tao and Li, Chuan-Feng and Xu, Jin-Shi and Guo, Guang-Can},
  journal={Physical Review Physics Education Research},
  volume={17},
  number={2},
  pages={020142},
  year={2021},
  publisher={APS},
  doi={10.1103/PhysRevPhysEducRes.17.020142}
}

@article{tu2023students,
  title={Students’ difficulties with the {Dirac} delta function in quantum mechanics},
  author={Tu, Tao and Li, Chuan-Feng and Xu, Jin-Shi and Guo, Guang-Can},
  journal={Physical Review Physics Education Research},
  volume={19},
  number={1},
  pages={010104},
  year={2023},
  publisher={APS},
  doi={10.1103/PhysRevPhysEducRes.19.010104}
}

@article{modir2019framing,
  title={Framing difficulties in quantum mechanics},
  author={Modir, Bahar and Thompson, John D and Sayre, Eleanor C},
  journal={Physical Review Physics Education Research},
  volume={15},
  number={2},
  pages={020146},
  year={2019},
  publisher={APS},
  doi={10.1103/PhysRevPhysEducRes.15.020146}
}

@article{modir2017students,
  title={Students’ epistemological framing in quantum mechanics problem solving},
  author={Modir, Bahar and Thompson, John D and Sayre, Eleanor C},
  journal={Physical Review Physics Education Research},
  volume={13},
  number={2},
  pages={020108},
  year={2017},
  publisher={APS},
  doi={10.1103/PhysRevPhysEducRes.13.020108}
}

@article{krijtenburg2017insights,
  title={Insights into teaching quantum mechanics in secondary and lower undergraduate education},
  author={Krijtenburg-Lewerissa, Kim and Pol, Hendrik Jan and Brinkman, Alexander and van Joolingen, Wouter R},
  journal={Physical Review Physics Education Research},
  volume={13},
  number={1},
  pages={010109},
  year={2017},
  publisher={APS},
  doi={10.1103/PhysRevPhysEducRes.13.010109}
}

@article{singh2015review,
  title={Review of student difficulties in upper-level quantum mechanics},
  author={Singh, Chandralekha and Marshman, Emily},
  journal={Physical Review Special Topics---Physics Education Research},
  volume={11},
  number={2},
  pages={020117},
  year={2015},
  publisher={APS},
  doi={10.1103/PhysRevSTPER.11.020117}
}

@article{marshman2017investigating,
  title={Investigating and improving student understanding of quantum mechanics in the context of single photon interference},
  author={Marshman, Emily and Singh, Chandralekha},
  journal={Physical Review Physics Education Research},
  volume={13},
  number={1},
  pages={010117},
  year={2017},
  publisher={APS},
  doi={10.1103/PhysRevPhysEducRes.13.010117}
}

@article{marshman2019validation,
  title={Validation and administration of a conceptual survey on the formalism and postulates of quantum mechanics},
  author={Marshman, Emily and Singh, Chandralekha},
  journal={Physical Review Physics Education Research},
  volume={15},
  number={2},
  pages={020128},
  year={2019},
  publisher={APS},
  doi={10.1103/PhysRevPhysEducRes.15.020128}
}

@article{sayer2017quantum,
  title={Quantum interactive learning tutorial on the double-slit experiment to improve student understanding of quantum mechanics},
  author={Sayer, Ryan and Maries, Alexandru and Singh, Chandralekha},
  journal={Physical Review Physics Education Research},
  volume={13},
  number={1},
  pages={010123},
  year={2017},
  publisher={APS},
  doi={10.1103/PhysRevPhysEducRes.13.010123}
}

@article{keebaugh2019improving,
  title={Improving student understanding of corrections to the energy spectrum of the hydrogen atom for the {Zeeman} effect},
  author={Keebaugh, Christof and Marshman, Emily and Singh, Chandralekha},
  journal={Physical Review Physics Education Research},
  volume={15},
  number={1},
  pages={010113},
  year={2019},
  publisher={APS},
  doi={10.1103/PhysRevPhysEducRes.15.010113}
}

@article{cook2006visual,
  title={Visual representations in science education: The influence of prior knowledge and cognitive load theory on instructional design principles},
  author={Cook, Michelle Patrick},
  journal={Science Education},
  volume={90},
  number={6},
  pages={1073--1091},
  year={2006},
  publisher={Wiley Online Library},
  doi={10.1002/sce.20164}
}

@article{soeharto2019review,
  title={A review of students’ common misconceptions in science and their diagnostic assessment tools},
  author={Soeharto, Soeharto and Csap{\'o}, Ben{\H{o}} and Sarimanah, Eri and Dewi, FI and Sabri, Tahmid},
  journal={Jurnal Pendidikan IPA Indonesia},
  volume={8},
  number={2},
  pages={247--266},
  year={2019},
  doi={10.15294/jpii.v8i2.18649}
}

@article{hasan1999misconceptions,
  title={Misconceptions and the certainty of response index ({CRI})},
  author={Hasan, Saleem and Bagayoko, Diola and Kelley, Ella L},
  journal={Physics Education},
  volume={34},
  number={5},
  pages={294--299},
  year={1999},
  publisher={Institute of Physics},
  doi={10.1088/0031-9120/34/5/304}
}

@article{gilbert1983concepts,
  title={Concepts, misconceptions and alternative conceptions: Changing perspectives in science education},
  author={Gilbert, John K and Watts, D Michael},
  year={1983},
  journal={Studies in Science Education},
  doi={10.1080/03057268308559905},
  volume={10},
  number={1},
  pages={61--98}
}

@article{kulgemeyer2023misconceptions,
  title={Misconceptions in physics explainer videos and the illusion of understanding: An experimental study},
  author={Kulgemeyer, Christoph and Wittwer, J{\"o}rg},
  journal={International Journal of Science and Mathematics Education},
  volume={21},
  number={2},
  pages={417--437},
  year={2023},
  publisher={Springer},
  doi={10.1007/s10763-022-10265-7}
}

@article{suprapto2020we,
  title={Do we experience misconceptions?: An ontological review of misconceptions in science},
  author={Suprapto, Nadi},
  journal={Studies in Philosophy of Science and Education},
  volume={1},
  number={2},
  pages={50--55},
  year={2020}
}

@article{resbiantoro2022review,
  title={A review of misconception in physics: the diagnosis, causes, and remediation},
  author={Resbiantoro, Gaguk and Setiani, Rahyu and {Dwikoranto}},
  journal={Journal of Turkish Science Education},
  volume={19},
  number={2},
  year={2022},
  doi={10.36681/tused.2022.128},
  pages={403--427}
}

@inproceedings{muller2007tackling,
  title={Tackling misconceptions in introductory physics using multimedia presentations},
  author={Muller, Derek A and Sharma, Manjula D},
  booktitle={UniServe Science Teaching and Learning Research Proceedings},
  year={2007},
  pages={58--63}
}

@article{miller2013role,
  title={Role of physics lecture demonstrations in conceptual learning},
  author={Miller, Kelly and Lasry, Nathaniel and Chu, Kelvin and Mazur, Eric},
  journal={Physical Review Special Topics---Physics Education Research},
  volume={9},
  number={2},
  pages={020113},
  year={2013},
  publisher={APS},
  doi={10.1103/PhysRevSTPER.9.020113}
}

@article{myneni2013interactive,
  title={An interactive and intelligent learning system for physics education},
  author={Myneni, Lakshman S and Narayanan, N Hari and Rebello, Sanjay and Rouinfar, Amy and Pumtambekar, Sadhana},
  journal={IEEE Transactions on Learning Technologies},
  volume={6},
  number={3},
  pages={228--239},
  year={2013},
  publisher={IEEE},
  doi={10.1109/TLT.2013.26}
}

@article{leinonen2013overcoming,
  title={Overcoming students’ misconceptions concerning thermal physics with the aid of hints and peer interaction during a lecture course},
  author={Leinonen, Risto and Asikainen, Mervi A and Hirvonen, Pekka E},
  journal={Physical Review Special Topics---Physics Education Research},
  volume={9},
  number={2},
  pages={020112},
  year={2013},
  publisher={APS},
  doi={10.1103/PhysRevSTPER.9.020112}
}

@article{hein1999using,
  title={Using writing to confront student misconceptions in physics},
  author={Hein, Teresa L},
  journal={European Journal of Physics},
  volume={20},
  number={3},
  pages={137--141},
  year={1999},
  publisher={IOP Publishing},
  doi={10.1088/0143-0807/20/3/002}
}

@article{liu2016student,
  title={Student misconceptions about force and acceleration in physics and engineering mechanics education},
  author={Liu, Gang and Fang, Ning},
  journal={International Journal of Engineering Education},
  volume={32},
  number={1},
  pages={19--29},
  year={2016}
}

@article{yuberti2020four,
  title={Four-tier diagnostic test with certainty of response index to identify misconception in physics},
  author={Yuberti, Yuberti and Suryani, Yani and Kurniawati, Indah},
  journal={Indonesian Journal of Science and Mathematics Education},
  volume={3},
  number={2},
  pages={245--253},
  year={2020},
  doi={10.24042/ijsme.v3i2.6061}
}

@inproceedings{diani2019four,
  title={Four-tier diagnostic test with certainty of response index on the concepts of fluid},
  author={Diani, Rahma and Alfin, J and Anggraeni, YM and Mustari, M and Fujiani, D},
  booktitle={Journal of Physics: Conference Series},
  volume={1155},
  pages={012078},
  year={2019},
  organization={IOP Publishing},
  doi={10.1088/1742-6596/1155/1/012078}
}

@inproceedings{fadllan2019analysis,
  title={Analysis of students’ misconceptions on mechanics using three-tier diagnostic test and clinical interview},
  author={Fadllan, A. and Prawira, W. Y. and {Arsini} and {Hartono}},
  booktitle={Journal of Physics: Conference Series},
  volume={1170},
  pages={012027},
  year={2019},
  organization={IOP Publishing},
  doi={10.1088/1742-6596/1170/1/012027}
}

@article{saglam2015confidence,
  title={The Confidence-Accuracy Relationship in Diagnostic Assessment: The Case of the Potential Difference in Parallel Electric Circuits.},
  author={Saglam, Murat},
  journal={Educational Sciences: Theory and Practice},
  volume={15},
  number={6},
  pages={1575--1584},
  year={2015},
  doi={10.12738/estp.2016.1.0033}
}

@article{gurel2015review,
  title={A review and comparison of diagnostic instruments to identify students' misconceptions in science},
  author={Kaltakci Gurel, Derya and Ery{\i}lmaz, Ali and McDermott, Lillian Christie},
  year={2015},
  doi={10.12973/eurasia.2015.1369a},
  journal={Eurasia Journal of Mathematics, Science and Technology Education},
  volume={11},
  number={5},
  pages={989--1008}
}

@article{shabani2010vygotsky,
  title={Vygotsky's zone of proximal development: Instructional implications and teachers' professional development.},
  author={Shabani, Karim and Khatib, Mohamad and Ebadi, Saman},
  journal={English Language Teaching},
  volume={3},
  number={4},
  pages={237--248},
  year={2010},
  publisher={Canadian Center of Science and Education},
  doi={10.5539/elt.v3n4p237}
}

@book{national1997science,
  title={Science teaching reconsidered: A handbook},
  author={{National Research Council}},
  year={1997},
  publisher={National Academies Press},
  doi={10.17226/5287}
}

@article{djanette2014determination,
  title={Determination of university students’ misconceptions about light using concept maps},
  author={Blizak, Djanette and Chafiqi, Fouad},
  journal={Procedia---Social and Behavioral Sciences},
  volume={152},
  pages={582--589},
  year={2014},
  publisher={Elsevier},
  note={Conference proceedings article},
  doi={10.1016/j.sbspro.2014.09.247}
}

@article{ogan2007effects,
  title={Effects of model-based teaching on pre-service physics teachers’ conceptions of the moon, moon phases, and other lunar phenomena},
  author={Ogan-Bekiroglu, Feral},
  journal={International Journal of Science Education},
  volume={29},
  number={5},
  pages={555--593},
  year={2007},
  publisher={Taylor \& Francis},
  doi={10.1080/09500690600718104}
}

@article{baser2007effectiveness,
  title={Effectiveness of conceptual change instruction on understanding of heat and temperature concepts},
  author={Baser, Mustafa and Geban, {\"O}mer},
  journal={Research in Science \& Technological Education},
  volume={25},
  number={1},
  pages={115--133},
  year={2007},
  publisher={Taylor \& Francis},
  doi={10.1080/02635140601053690}
}

@article{csahin2010computer,
  title={Computer supported conceptual change text: Fluid pressure},
  author={{\c{S}}ahin, {\c{C}}i{\u{g}}dem and {\.I}pek, Hava and {\c{C}}epni, Salih},
  journal={Procedia---Social and Behavioral Sciences},
  volume={2},
  number={2},
  pages={922--927},
  year={2010},
  publisher={Elsevier},
  doi={10.1016/j.sbspro.2010.03.127},
  note={Conference proceedings article}
}

@article{franco2012examining,
  title={Examining the influences of epistemic beliefs and knowledge representations on cognitive processing and conceptual change when learning physics},
  author={Franco, Gina M and Muis, Krista R and Kendeou, Panayiota and Ranellucci, John and Sampasivam, Lavanya and Wang, Xihui},
  journal={Learning and Instruction},
  volume={22},
  number={1},
  pages={62--77},
  year={2012},
  publisher={Elsevier},
  doi={10.1016/j.learninstruc.2011.06.003}
}

@article{durmucs2010effects,
  title={Effects of conceptual change texts and laboratory experiments on fourth grade students’ understanding of matter and change concepts},
  author={Durmu{\c{s}}, Jale and Bayraktar, {\c{S}}ule},
  journal={Journal of Science Education and Technology},
  volume={19},
  pages={498--504},
  year={2010},
  publisher={Springer},
  doi={10.1007/s10956-010-9216-9}
}

@article{wendt2014effect,
  title={The effect of online collaboration on middle school student science misconceptions as an aspect of science literacy},
  author={Wendt, Jillian L and Rockinson-Szapkiw, Amanda},
  journal={Journal of Research in Science Teaching},
  volume={51},
  number={9},
  pages={1103--1118},
  year={2014},
  publisher={Wiley Online Library},
  doi={10.1002/tea.21169}
}

@article{korganci2015importance,
  title={The importance of inquiry-based learning on electric circuit models for conceptual understanding},
  author={Korganci, Nuri and Miron, Cristina and Dafinei, Adrian and Antohe, Stefan},
  journal={Procedia---Social and Behavioral Sciences},
  volume={191},
  pages={2463--2468},
  year={2015},
  publisher={Elsevier},
  doi={10.1016/j.sbspro.2015.04.530},
  note={Conference proceedings article}
}

@article{dutt2012decisions,
  title={Decisions from experience reduce misconceptions about climate change},
  author={Dutt, Varun and Gonzalez, Cleotilde},
  journal={Journal of Environmental Psychology},
  volume={32},
  number={1},
  pages={19--29},
  year={2012},
  publisher={Elsevier},
  doi={10.1016/j.jenvp.2011.10.003}
}

@article{falloon2019using,
  title={Using simulations to teach young students science concepts: An Experiential Learning theoretical analysis},
  author={Falloon, Garry},
  journal={Computers \& Education},
  volume={135},
  pages={138--159},
  year={2019},
  publisher={Elsevier},
  doi={10.1016/j.compedu.2019.03.001}
}

@article{hockicko2014correcting,
  title={Correcting students’ misconceptions about automobile braking distances and video analysis using interactive program tracker},
  author={Hockicko, Peter and Trpi{\v{s}}ov{\'a}, Be{\'a}ta and Ondru{\v{s}}, J{\'a}n},
  journal={Journal of Science Education and Technology},
  volume={23},
  pages={763--776},
  year={2014},
  publisher={Springer},
  doi={10.1007/s10956-014-9510-z}
}

@article{kozhevnikov2013learning,
  title={Learning relative motion concepts in immersive and non-immersive virtual environments},
  author={Kozhevnikov, Michael and Gurlitt, Johannes and Kozhevnikov, Maria},
  journal={Journal of Science Education and Technology},
  volume={22},
  pages={952--962},
  year={2013},
  publisher={Springer},
  doi={10.1007/s10956-013-9441-0}
}

@article{phanphech2019explaining,
  title={Explaining and enacting for conceptual understanding in secondary school physics},
  author={Phanphech, Parinda and Tanitteerapan, Tanes and Murphy, Elizabeth},
  journal={Issues in Educational Research},
  volume={29},
  number={1},
  pages={180--204},
  year={2019}
}

@article{schneps2014conceptualizing,
  title={Conceptualizing astronomical scale: Virtual simulations on handheld tablet computers reverse misconceptions},
  author={Schneps, Matthew H and Ruel, Jonathan and Sonnert, Gerhard and Dussault, Mary and Griffin, Michael and Sadler, Philip M},
  journal={Computers \& Education},
  volume={70},
  pages={269--280},
  year={2014},
  publisher={Elsevier},
  doi={10.1016/j.compedu.2013.09.001}
}

@article{smith1994misconceptions,
  title={Misconceptions reconceived: A constructivist analysis of knowledge in transition},
  author={Smith, III, John P. and diSessa, Andrea A. and Roschelle, Jeremy},
  journal={Journal of the Learning Sciences},
  volume={3},
  number={2},
  pages={115--163},
  year={1993},
  doi={10.1207/s15327809jls0302_1}
}

@incollection{scott1991teaching,
  title={Teaching for conceptual change: A review of strategies},
  author={Scott, P. H. and Asoko, H. M. and Driver, R. H.},
  year={1998},
  publisher={International Commission on Physics Education},
  booktitle={Connecting Research in Physics Education with Teacher Education},
  editor={Tiberghien, Andr{\'e}e and Jossem, E. Leonard and Barojas, Jorge},
  volume={1},
  chapter={C5},
  url={https://zenodo.org/records/6463545},
  note={Reprinted from an IPN workshop contribution (March 1991)}
}

@article{styer1996common,
  title={Common misconceptions regarding quantum mechanics},
  author={Styer, Daniel F},
  journal={American Journal of Physics},
  volume={64},
  number={1},
  pages={31--34},
  year={1996},
  doi={10.1119/1.18288}
}

@article{busetto2020use,
  title={How to use and assess qualitative research methods},
  author={Busetto, Loraine and Wick, Wolfgang and Gumbinger, Christoph},
  journal={Neurological Research and Practice},
  volume={2},
  number={1},
  pages={14},
  year={2020},
  publisher={Springer},
  doi={10.1186/s42466-020-00059-z}
}

@article{dilley2000conducting,
  title={Conducting successful interviews: Tips for intrepid research},
  author={Dilley, Patrick},
  journal={Theory Into Practice},
  volume={39},
  number={3},
  pages={131--137},
  year={2000},
  publisher={Taylor \& Francis},
  doi={10.1207/s15430421tip3903_3}
}

@article{brod2009qualitative,
  title={Qualitative research and content validity: developing best practices based on science and experience},
  author={Brod, Meryl and Tesler, Laura E and Christensen, Torsten L},
  journal={Quality of Life Research},
  volume={18},
  pages={1263--1278},
  year={2009},
  publisher={Springer},
  doi={10.1007/s11136-009-9540-9}
}

@article{solarino2021challenges,
  title={Challenges and best-practice recommendations for designing and conducting interviews with elite informants},
  author={Solarino, Angelo M and Aguinis, Herman},
  journal={Journal of Management Studies},
  volume={58},
  number={3},
  pages={649--672},
  year={2021},
  publisher={Wiley Online Library},
  doi={10.1111/joms.12620}
}

@article{asfaw2022building,
  title={Building a quantum engineering undergraduate program},
  author={Asfaw, Abraham and Blais, Alexandre and Brown, Kenneth R and Candelaria, Jonathan and Cantwell, Christopher and Carr, Lincoln D and Combes, Joshua and Debroy, Dripto M and Donohue, John M and Economou, Sophia E and others},
  journal={IEEE Transactions on Education},
  volume={65},
  number={2},
  pages={220--242},
  year={2022},
  publisher={IEEE},
  doi={10.1109/TE.2022.3144943}
}

@article{ WOS:A1996WE28100002,
Author = {Magara, K},
Title = {Why is it difficult to rectify a misconception in learners?},
Journal = {Japanese Journal of Educational Psychology},
Year = {1996},
Volume = {44},
Number = {4},
Pages = {379-388},
Month = {DEC},
DOI = {10.5926/jjep1953.44.4\_379},
ISSN = {0021-5015},
Unique-ID = {WOS:A1996WE28100002},
}

@article{ WOS:001088164200001,
Author = {Ozmen, Kubra},
Title = {Health Science Students' Conceptual Understanding of Electricity:
   Misconception or Lack of Knowledge?},
Journal = {Research in Science Education},
Year = {2024},
Volume = {54},
Number = {2},
Pages = {225-243},
Month = {APR},
DOI = {10.1007/s11165-023-10136-3},
EarlyAccessDate = {OCT 2023},
ISSN = {0157-244X},
EISSN = {1573-1898},
ResearcherID-Numbers = {Özmen, Kübra/AAD-2638-2020
   },
ORCID-Numbers = {Ozmen, Kubra/0000-0001-7838-8314},
Unique-ID = {WOS:001088164200001},
}

@article{ WOS:000943130300001,
Author = {Lin, Jiabei and Xing, Yuting and Hu, Yudi and Zhang, Jian and Bao, Lei
   and Luo, Kaiqing and Yu, Keke and Xiao, Yang},
Title = {Inhibitory control involvement in overcoming the position-velocity
   indiscrimination misconception among college physics majors},
Journal = {Physical Review Physics Education Research},
Year = {2023},
Volume = {19},
Number = {1},
Month = {FEB 23},
DOI = {10.1103/PhysRevPhysEducRes.19.010112},
Article-Number = {010112},
ISSN = {2469-9896},
ResearcherID-Numbers = {Luo, Kaiqing/ABD-6166-2020
   Xiao, Yang/AAE-6592-2022
   },
ORCID-Numbers = {Luo, Kaiqing/0000-0002-6278-0917
   Yu, Keke/0000-0002-8210-0974
   Xiao, Yang/0000-0002-2571-7759},
Unique-ID = {WOS:000943130300001},
  eid={010112}
}

@article{ WOS:000608978200006,
Author = {Suryadi, A. and Kusairi, S. and Husna, D. A.},
Title = {Comparative Study of Secondary School Students' and Pre-Service
   Teachers' Misconception about Simple Electric Circuit},
Journal = {Jurnal Pendidikan Fisika Indonesia},
Year = {2020},
Volume = {16},
Number = {2},
Pages = {111-121},
Month = {DEC},
DOI = {10.15294/jpfi.v16i2.21909},
ISSN = {1693-1246},
EISSN = {2355-3812},
ResearcherID-Numbers = {suryadi, Ahmad/AAX-7068-2021
   },
ORCID-Numbers = {Suryadi, Ahmad/0000-0003-2375-916X},
Unique-ID = {WOS:000608978200006},
}

@article{ WOS:000245278400006,
Author = {Chang, Huey-Por and Chen, Jun-Yi and Guo, Chorng-Jee and Chen,
   Chung-Chih and Chang, Ching-Yi and Lin, Shean-Huei and Su, Wei-Jou and
   Lain, Kuen-Der and Hsu, Shun-Yi and Lin, Jang-Long and Chen, Chin-Chang
   and Cheng, Yi-Ting and Wang, Loung-Shyi and Tseng, Yaw-Teng},
Title = {Investigating primary and secondary students' learning of physics
   concepts in {Taiwan}},
Journal = {International Journal of Science Education},
Year = {2007},
Volume = {29},
Number = {4},
Pages = {465-482},
DOI = {10.1080/09500690601073210},
ISSN = {0950-0693},
ResearcherID-Numbers = {Hung, Chun-Hsiung/D-5521-2009
   Chen, Chien/Q-3826-2018
   Chang, Chun-Hung/AAT-1641-2021
   Chen, Chien-Chin/A-9977-2016},
Unique-ID = {WOS:000245278400006},
}

@article{ WOS:000298904800024,
Author = {Kizilcik, Hasan Sahin and Gunes, Bilal},
Title = {DEVELOPING THREE-TIRE MISCONCEPTION TEST ABOUT REGULAR CIRCULAR MOTION},
Journal = {Hacettepe University Journal of Education},
Year = {2011},
Number = {41},
Pages = {278-292},
ISSN = {1300-5340},
ResearcherID-Numbers = {GUNES, Prof. Dr. Bilal/A-4520-2017
   Kizilcik, Hasan Sahin/A-7692-2018},
ORCID-Numbers = {GUNES, Prof. Dr. Bilal/0000-0002-3911-6778
   Kizilcik, Hasan Sahin/0000-0001-8622-0765},
Unique-ID = {WOS:000298904800024},
}

@article{ WOS:000642957600001,
Author = {Halim, Abdul and Mahzum, Elmi and Yacob, Muhammad and Irwandi, Irwandi
   and Halim, Lilia},
Title = {The Impact of Narrative Feedback, E-Learning Modules and Realistic Video
   and the Reduction of Misconception},
Journal = {Education Sciences},
Year = {2021},
Volume = {11},
Number = {4},
Month = {APR},
DOI = {10.3390/educsci11040158},
Article-Number = {158},
EISSN = {2227-7102},
ResearcherID-Numbers = {Halim, Lilia/AAK-8167-2020
   Halim, A/K-1225-2015
   Irwandi, Irwandi/GPS-8171-2022
   Mahzum, Elmi/P-9636-2017
   Halim, M.Si, Prof. Dr. A./E-9707-2017},
ORCID-Numbers = {Mahzum, Elmi/0000-0003-0359-5110
   Irwandi, Irwandi/0000-0001-5263-8338
   Halim, M.Si, Prof. Dr. A./0000-0001-7140-6161},
Unique-ID = {WOS:000642957600001},
  eid={158}
}

@article{ WOS:001129334900001,
Author = {Ding, Lu and Li, Tong and Jiang, Shiyan and Gapud, Albert},
Title = {Students' perceptions of using {ChatGPT} in a physics class as a virtual
   tutor},
Journal = {International Journal of Educational Technology in Higher Education},
Year = {2023},
Volume = {20},
Number = {1},
Month = {DEC 22},
DOI = {10.1186/s41239-023-00434-1},
Article-Number = {63},
ISSN = {2365-9440},
ResearcherID-Numbers = {Ding, Lu/JNE-3772-2023
   Li, Tong/JXO-0453-2024
   },
ORCID-Numbers = {Ding, Lu/0000-0002-1454-1532
   Li, Tong/0000-0002-1166-625X},
Unique-ID = {WOS:001129334900001},
  eid={63}
}

@article{ WOS:000181045900006,
Author = {Prange, HD},
Title = {{Laplace}'s law and the alveolus: A misconception of anatomy and a
   misapplication of physics},
Journal = {Advances in Physiology Education},
Year = {2003},
Volume = {27},
Number = {1},
Pages = {34-40},
Month = {MAR},
DOI = {10.1152/advan.00024.2002},
ISSN = {1043-4046},
EISSN = {1522-1229},
Unique-ID = {WOS:000181045900006},
}

@article{ WOS:000483008600010,
Author = {Weingartner, Kristin M. and Masnick, Amy M.},
Title = {Refutation texts: Implying the refutation of a scientific misconception
   can facilitate knowledge revision},
Journal = {Contemporary Educational Psychology},
Year = {2019},
Volume = {58},
Pages = {138-148},
Month = {JUL},
DOI = {10.1016/j.cedpsych.2019.03.004},
ISSN = {0361-476X},
EISSN = {1090-2384},
Unique-ID = {WOS:000483008600010},
}

@article{ WOS:000283554300001,
Author = {Chiou, Guo-Li and Anderson, O. Roger},
Title = {A Multi-dimensional Cognitive Analysis of Undergraduate Physics
   Students' Understanding of Heat Conduction},
Journal = {International Journal of Science Education},
Year = {2010},
Volume = {32},
Number = {16},
Pages = {2113-2142},
DOI = {10.1080/09500690903258246},
Article-Number = {PII 916259467},
ISSN = {0950-0693},
Unique-ID = {WOS:000283554300001},
}

@article{ WOS:000265992600004,
Author = {Ozdemir, Omer Faruk},
Title = {Avoidance from Thought Experiments: Fear of misconception},
Journal = {International Journal of Science Education},
Year = {2009},
Volume = {31},
Number = {8},
Pages = {1049-1068},
DOI = {10.1080/09500690801932538},
Article-Number = {PII 791265276},
ISSN = {0950-0693},
EISSN = {1464-5289},
ResearcherID-Numbers = {ozdemir, omer faruk/P-9253-2016},
ORCID-Numbers = {ozdemir, omer faruk/0000-0003-4715-195X},
Unique-ID = {WOS:000265992600004},
}

@article{ WOS:000340479200002,
Author = {Wong, Chee Leong and Chu, Hye-Eun and Yap, Kueh Chin},
Title = {Developing a Framework for Analyzing Definitions: A study of {{\emph{The Feynman Lectures}}}},
Journal = {International Journal of Science Education},
Year = {2014},
Volume = {36},
Number = {15},
Pages = {2481-2513},
DOI = {10.1080/09500693.2014.893594},
ISSN = {0950-0693},
EISSN = {1464-5289},
ORCID-Numbers = {chu, hye-eun/0000-0001-8937-1446},
Unique-ID = {WOS:000340479200002},
}

@article{ WOS:000836287700001,
Author = {Hull, Michael M. and Jansky, Alexandra and Hopf, Martin},
Title = {Does confidence in a wrong answer imply a misconception?},
Journal = {Physical Review Physics Education Research},
Year = {2022},
Volume = {18},
Number = {2},
Month = {AUG 2},
DOI = {10.1103/PhysRevPhysEducRes.18.020108},
Article-Number = {020108},
ISSN = {2469-9896},
ResearcherID-Numbers = {Jansky, Alexandra/LBH-4841-2024
   Hopf, Martin/IAM-3246-2023
   },
ORCID-Numbers = {Hopf, Martin/0000-0003-3273-0000
   Hull, Michael M./0000-0003-1551-9553
   Jansky, Alexandra/0000-0003-0594-2390},
Unique-ID = {WOS:000836287700001},
  eid={020108}
}

@article{ WOS:000265896800003,
Author = {Cavallini, Graziano and Giliberti, Marco},
Title = {La lezione della fisica quantistica},
Journal = {Epistemologia},
Year = {2008},
Volume = {31},
Number = {2},
Pages = {219--240},
Month = {JUL-DEC},
ISSN = {0392-9760},
ResearcherID-Numbers = {Giliberti, Marco/AAE-3258-2022},
Unique-ID = {WOS:000265896800003},
}

@article{ WOS:000269662300015,
Author = {Halim, A. and Meerah, T. Subahan and Halim, Lilia},
Title = {Development and Application of Diagnostic Test to Identify Students'
   Misconceptions of Quantum Physics},
Journal = {Sains Malaysiana},
Year = {2009},
Volume = {38},
Number = {4},
Pages = {543-551},
Month = {AUG},
ISSN = {0126-6039},
ResearcherID-Numbers = {Halim, Lilia/AAK-8167-2020
   Halim, A/K-1225-2015
   Halim, M.Si, Prof. Dr. A./E-9707-2017},
ORCID-Numbers = {Halim, M.Si, Prof. Dr. A./0000-0001-7140-6161},
Unique-ID = {WOS:000269662300015},
}

@article{ WOS:000533707200001,
Author = {Hull, Michael M. and Jansky, Alexandra and Hopf, Martin},
Title = {Probability-related naive ideas across physics topics},
Journal = {Studies in Science Education},
Year = {2021},
Volume = {57},
Number = {1},
Pages = {45-83},
Month = {JAN 2},
DOI = {10.1080/03057267.2020.1757244},
EarlyAccessDate = {MAY 2020},
ISSN = {0305-7267},
EISSN = {1940-8412},
ResearcherID-Numbers = {Hopf, Martin/IAM-3246-2023
   Jansky, Alexandra/LBH-4841-2024
   },
ORCID-Numbers = {Hull, Michael M./0000-0003-1551-9553
   Hopf, Martin/0000-0003-3273-0000
   Jansky, Alexandra/0000-0003-0594-2390},
Unique-ID = {WOS:000533707200001},
}

@article{ WOS:000357483800002,
Author = {Mou, Yi and Zhu, Liqi and Chen, Zhe},
Title = {Developmental changes in children's understanding of horizontal
   projectile motion},
Journal = {International Journal of Psychology},
Year = {2015},
Volume = {50},
Number = {4},
Pages = {256-264},
Month = {AUG},
DOI = {10.1002/ijop.12095},
ISSN = {0020-7594},
EISSN = {1464-066X},
ResearcherID-Numbers = {Mou, yikun/GYK-1908-2022},
Unique-ID = {WOS:000357483800002},
}

@article{ WOS:000285629600004,
Author = {Eryilmaz, Ali},
Title = {Development and Application of Three-Tier Heat and Temperature Test:
   Sample of Bachelor and Graduate Students},
Journal = {Eurasian Journal of Educational Research},
Year = {2010},
Volume = {10},
Number = {40},
Pages = {53-76},
Month = {SUM},
ISSN = {1302-597X},
EISSN = {2528-8911},
ResearcherID-Numbers = {ERYILMAZ, ALI/C-3398-2013},
ORCID-Numbers = {ERYILMAZ, ALI/0000-0003-2161-6018},
Unique-ID = {WOS:000285629600004},
}

@article{ WOS:001167417300002,
Author = {Shaafi, N. F. and Yusof, M. M. M. and Ellianawati, E. and Aziz, S. N. A.},
Title = {A Study of Interest in Astronomy Among University Students in {Malaysia}},
Journal = {Jurnal Pendidikan Fisika Indonesia},
Year = {2023},
Volume = {19},
Number = {2},
Pages = {108-121},
Month = {DEC},
DOI = {10.15294/jpfi.v19i2.44240},
ISSN = {1693-1246},
EISSN = {2355-3812},
ResearcherID-Numbers = {Mohd Yusof, Mohammad Mubarrak/GPW-8792-2022
   },
ORCID-Numbers = {MOHD YUSOF, MOHAMMAD MUBARRAK/0000-0003-2742-6629
   SHAAFI, NUR FARHA/0000-0003-4511-2746},
Unique-ID = {WOS:001167417300002},
}

@article{ WOS:000334106800003,
Author = {Ergin, Serap and Atasoy, Sengul},
Title = {COMPARATIVE ANALYSIS OF THE EFFECTIVENESS OF {4MAT} TEACHING METHOD IN
   REMOVING PUPILS' PHYSICS MISCONCEPTIONS OF ELECTRICITY},
Journal = {Journal of Baltic Science Education},
Year = {2013},
Volume = {12},
Number = {6},
Pages = {730-746},
ISSN = {1648-3898},
ResearcherID-Numbers = {Atasoy, Şengül/AGQ-2563-2022
   },
ORCID-Numbers = {Atasoy, Sengul/0000-0002-7664-1010},
Unique-ID = {WOS:000334106800003},
}

@article{ WOS:000335541700018,
Author = {Acar, Omer},
Title = {Scientific reasoning, conceptual knowledge, \& achievement differences
   between prospective science teachers having a consistent misconception
   and those having a scientific conception in an argumentation-based
   guided inquiry course},
Journal = {Learning and Individual Differences},
Year = {2014},
Volume = {30},
Pages = {148-154},
Month = {FEB},
DOI = {10.1016/j.lindif.2013.12.002},
ISSN = {1041-6080},
EISSN = {1873-3425},
ResearcherID-Numbers = {Acar, Omer/F-5091-2018},
ORCID-Numbers = {Acar, Omer/0000-0002-9369-5539},
Unique-ID = {WOS:000335541700018},
}

@incollection{ WOS:000568612900006,
Author = {Neidorf, Teresa and Arora, Alka and Erberber, Ebru and Tsokodayi,
   Yemurai and Mai, Thanh},
Book-Author = {Neidorf, T
   Arora, A
   Erberber, E
   Tsokodayi, Y
   Mai, T},
Title = {Conclusions About Using {TIMSS} and {TIMSS} Advanced Data to Explore Student
   Misconceptions, Errors, and Misunderstandings in Physics and Mathematics},
Booktitle = {Student Misconceptions and Errors in Physics and Mathematics: Exploring
   Data from TIMSS and TIMSS Advanced},
Series = {IEA Research for Education},
Year = {2020},
Volume = {9},
Pages = {133-153},
DOI = {10.1007/978-3-030-30188-0_5},
ISSN = {2366-1631},
EISSN = {2366-164X},
ISBN = {978-3-030-30188-0; 978-3-030-30187-3},
Unique-ID = {WOS:000568612900006},
Publisher = {Springer},
Address = {Cham},
Chapter = {5},
}

@article{ WOS:A1997XU39100003,
Author = {Guzzetti, BJ and Williams, WO and Skeels, SA and Wu, SM},
Title = {Influence of text structure on learning counterintuitive physics
   concepts},
Journal = {Journal of Research in Science Teaching},
Year = {1997},
Volume = {34},
Number = {7},
Pages = {701-719},
Month = {SEP},
DOI = {10.1002/(SICI)1098-2736(199709)34:7<701::AID-TEA3>3.0.CO;2-Q},
ISSN = {0022-4308},
EISSN = {1098-2736},
Unique-ID = {WOS:A1997XU39100003},
}

@article{ WOS:000304504700042,
Author = {Ince, Elif and Sesen, Burcin Acar and Kirbaslar, F. Gulay},
Title = {Investigation of undergraduate students' understanding of radiation and
   radioactivity},
Journal = {Energy Education Science and Technology Part B-Social and Educational
   Studies},
Year = {2012},
Volume = {4},
Number = {2},
Pages = {993-1004},
Month = {APR},
ISSN = {1308-7711},
ResearcherID-Numbers = {Kırbaşlar, Fatma/AAH-2389-2019
   Acar Sesen, Burcin/V-4542-2017
   Ince, Elif/R-1127-2018},
ORCID-Numbers = {Kirbaslar, Fatma Gulay/0000-0002-0267-9630
   Acar Sesen, Burcin/0000-0002-1585-0441
   Ince, Elif/0000-0003-4821-5441},
Unique-ID = {WOS:000304504700042},
}

@article{ WOS:000744253000025,
Author = {Soeharto, Soeharto and Csapo, Beno},
Title = {Evaluating item difficulty patterns for assessing student misconceptions
   in science across physics, chemistry, and biology concepts},
Journal = {Heliyon},
Year = {2021},
Volume = {7},
Number = {11},
Month = {NOV},
DOI = {10.1016/j.heliyon.2021.e08352},
EarlyAccessDate = {NOV 2021},
Article-Number = {e08352},
EISSN = {2405-8440},
ResearcherID-Numbers = {Soeharto, Soeharto/AAV-3876-2020
   Csapó, Benő/D-5634-2013
   },
ORCID-Numbers = {Soeharto, Soeharto/0000-0003-4332-7401},
Unique-ID = {WOS:000744253000025},
  eid={e08352}
}

@article{ WOS:000384323500003,
Author = {Motlhabane, Abraham},
Title = {LEARNER'S ALTERNATIVE AND MISCONCEPTIONS IN PHYSICS: A PHENOMENOGRAPHIC
   STUDY},
Journal = {Journal of Baltic Science Education},
Year = {2016},
Volume = {15},
Number = {4},
Pages = {424-440},
ISSN = {1648-3898},
ORCID-Numbers = {Motlhabane, Abraham/0000-0002-3874-4959},
Unique-ID = {WOS:000384323500003},
DOI = {10.33225/jbse/16.15.424},
}

@article{ WOS:000687380800003,
Author = {Abdullah, Mohd Nor Syahrir and Karpudewan, Mageswary and Tanimale, Bala
   Murali},
Title = {Executive function of the brain and its influences on understanding of
   physics concept},
Journal = {Trends in Neuroscience and Education},
Year = {2021},
Volume = {24},
Month = {SEP},
DOI = {10.1016/j.tine.2021.100159},
EarlyAccessDate = {AUG 2021},
Article-Number = {100159},
ISSN = {2452-0837},
EISSN = {2211-9493},
ResearcherID-Numbers = {Karpudewan, Mageswary/E-2947-2012
   Abdullah, Mohd Nor Syahrir/B-8843-2019},
Unique-ID = {WOS:000687380800003},
Note = {See corrigendum, DOI: 10.1016/j.tine.2022.100177},
  eid={100159}
}

@article{ WOS:000075852700007,
Author = {Taber, KS},
Title = {The sharing-out of nuclear attraction: or `I can't think about physics
   in chemistry'},
Journal = {International Journal of Science Education},
Year = {1998},
Volume = {20},
Number = {8},
Pages = {1001-1014},
Month = {SEP-OCT},
DOI = {10.1080/0950069980200807},
ISSN = {0950-0693},
ResearcherID-Numbers = {Taber, Keith/A-6996-2011},
Unique-ID = {WOS:000075852700007},
}

@article{ WOS:000345549500012,
Author = {Kolcak, Demet Yolas and Mogol, Selma and Unsal, Yasin},
Title = {A Comparison of the Effects of Laboratory Method and Computer
   Simulations to Avoid Misconceptions in Physics Education},
Journal = {Egitim ve Bilim-Education and Science},
Year = {2014},
Volume = {39},
Number = {175},
Pages = {154-171},
Month = {OCT},
ISSN = {1300-1337},
ResearcherID-Numbers = {ÜNSAL, Yasin/ABH-7918-2020},
Unique-ID = {WOS:000345549500012},
DOI = {10.15390/EB.2014.2052},
}

@article{ WOS:000856603000002,
Author = {Taslidere, Erdal and Yildirim, Bilgeturk},
Title = {Effect of Conceptual Change-Oriented Instruction on Students' Conceptual
   Understanding and Attitudes Towards Simple Electricity},
Journal = {International Journal of Science and Mathematics Education},
Year = {2023},
Volume = {21},
Number = {5},
Pages = {1567-1589},
Month = {JUN},
DOI = {10.1007/s10763-022-10319-w},
EarlyAccessDate = {SEP 2022},
ISSN = {1571-0068},
EISSN = {1573-1774},
ResearcherID-Numbers = {TASLIDERE, ERDAL/ABI-3943-2020
   },
ORCID-Numbers = {TASLIDERE, Erdal/0000-0002-3158-2483},
Unique-ID = {WOS:000856603000002},
}

@article{ WOS:000312337600003,
Author = {Lemmer, Miriam},
Title = {Nature, Cause and Effect of Students' Intuitive Conceptions Regarding
   Changes in Velocity},
Journal = {International Journal of Science Education},
Year = {2013},
Volume = {35},
Number = {2},
Pages = {239-261},
DOI = {10.1080/09500693.2011.647110},
ISSN = {0950-0693},
EISSN = {1464-5289},
Unique-ID = {WOS:000312337600003},
}

@article{ WOS:000430180900002,
Author = {Eshach, Haim and Lin, Tzu-Chiang and Tsai, Chin-Chung},
Title = {Misconception of sound and conceptual change: A cross-sectional study on
   students' materialistic thinking of sound},
Journal = {Journal of Research in Science Teaching},
Year = {2018},
Volume = {55},
Number = {5},
Pages = {664-684},
Month = {MAY},
DOI = {10.1002/tea.21435},
ISSN = {0022-4308},
EISSN = {1098-2736},
ResearcherID-Numbers = {Tsai, Chin-Chung/E-5902-2010
   Lin, Tzu-Chiang/G-1410-2012},
ORCID-Numbers = {Tsai, Chin-Chung/0000-0001-7744-9971
   Lin, Tzu-Chiang/0000-0003-3842-3749},
Unique-ID = {WOS:000430180900002},
}

@article{ WOS:001109203600004,
Author = {Pepino, Ron A. and Mabile, Risley W.},
Title = {A Misconception Regarding the {Einstein} Equivalence Principle and a
   Possible Cure Using the {Twin Paradox}},
Journal = {The Physics Teacher},
Year = {2023},
Volume = {61},
Number = {2},
Pages = {118-121},
Month = {FEB},
DOI = {10.1119/5.0075153},
ISSN = {0031-921X},
EISSN = {1943-4928},
ORCID-Numbers = {Pepino, Ron/0000-0003-4160-0558},
Unique-ID = {WOS:001109203600004},
}

@article{ WOS:000400179000007,
Author = {Kaltakci-Gurel, Derya and Eryilmaz, Ali and McDermott, Lillian Christie},
Title = {Development and application of a four-tier test to assess pre-service
   physics teachers' misconceptions about geometrical optics},
Journal = {Research in Science \& Technological Education},
Year = {2017},
Volume = {35},
Number = {2},
Pages = {238-260},
Month = {JUN},
DOI = {10.1080/02635143.2017.1310094},
ISSN = {0263-5143},
EISSN = {1470-1138},
ResearcherID-Numbers = {ERYILMAZ, ALI/C-3398-2013
   Kaltakci-Gurel, Derya/F-5981-2018},
ORCID-Numbers = {ERYILMAZ, ALI/0000-0003-2161-6018
   Kaltakci-Gurel, Derya/0000-0003-3727-7516},
Unique-ID = {WOS:000400179000007},
}

@article{ WOS:A1992GX44200002,
Author = {Brown, David E.},
Title = {USING EXAMPLES AND ANALOGIES TO REMEDIATE MISCONCEPTIONS IN PHYSICS -
   FACTORS INFLUENCING CONCEPTUAL CHANGE},
Journal = {Journal of Research in Science Teaching},
Year = {1992},
Volume = {29},
Number = {1},
Pages = {17-34},
Month = {JAN},
DOI = {10.1002/tea.3660290104},
ISSN = {0022-4308},
Unique-ID = {WOS:A1992GX44200002},
}

@article{ WOS:000281064200006,
Author = {Stocklmayer, Susan},
Title = {Teaching Direct Current Theory Using a Field Model},
Journal = {International Journal of Science Education},
Year = {2010},
Volume = {32},
Number = {13},
Pages = {1801-1828},
DOI = {10.1080/09500690903575748},
Article-Number = {PII 921248375},
ISSN = {0950-0693},
EISSN = {1464-5289},
ORCID-Numbers = {Stocklmayer, Susan/0000-0002-2363-1345},
Unique-ID = {WOS:000281064200006},
}

@article{ WOS:000403832500001,
Author = {Jaaskelainen, Markku and Lagerkvist, Andreas},
Title = {Why do they not answer and do they really learn? A case study in
   analysing student response flows in introductory physics using an
   audience response system},
Journal = {European Journal of Physics},
Year = {2017},
Volume = {38},
Number = {4},
Month = {JUL},
DOI = {10.1088/1361-6404/aa73b5},
Article-Number = {045705},
ISSN = {0143-0807},
EISSN = {1361-6404},
ResearcherID-Numbers = {Jääskeläinen, Markku/AAI-2520-2019
   },
ORCID-Numbers = {Jaaskelainen, Markku/0000-0002-8049-2425},
Unique-ID = {WOS:000403832500001},
  eid={045705}
}

@article{ WOS:000324551800009,
Author = {Dega, Bekele Gashe and Kriek, Jeanne and Mogese, Temesgen Fereja},
Title = {Categorization of Alternative Conceptions in Electricity and Magnetism:
   the Case of {Ethiopian} Undergraduate Students},
Journal = {Research in Science Education},
Year = {2013},
Volume = {43},
Number = {5},
Pages = {1891-1915},
Month = {OCT},
DOI = {10.1007/s11165-012-9332-z},
ISSN = {0157-244X},
EISSN = {1573-1898},
ResearcherID-Numbers = {Kriek, Jeanne/E-1312-2015
   },
ORCID-Numbers = {Kriek, Jeanne/0000-0001-6248-4563
   Dega, Bekele Gashe/0000-0002-2115-1348},
Unique-ID = {WOS:000324551800009},
}

@article{ WOS:000291686800022,
Author = {Wong, Darren and Lee, Paul and Gao, Shenghan and Wang, Xuezhou and Huan, Yan Qi and Foong, See Kit},
Title = {The photoelectric effect: experimental confirmation concerning a
   widespread misconception in the theory},
Journal = {European Journal of Physics},
Year = {2011},
Volume = {32},
Number = {4},
Pages = {1059-1064},
Month = {JUL},
DOI = {10.1088/0143-0807/32/4/018},
ISSN = {0143-0807},
ResearcherID-Numbers = {Lee, Paul/A-8749-2008
   Foong, See/I-6713-2012
   },
ORCID-Numbers = {Wang, Aiden/0000-0001-6335-9836
   Wong, Darren/0009-0006-8870-5639
   Huan, Yan Qi/0000-0002-6505-7150},
Unique-ID = {WOS:000291686800022},
}

@article{ WOS:000318278800002,
Author = {Martinez-Borreguero, Guadalupe and P{\'e}rez-Rodr{\'i}guez, {\'A}ngel Luis and Suero-L{\'o}pez, Mar{\'i}a Isabel and Pardo-Fern{\'a}ndez, Pedro Jos{\'e}},
Title = {Detection of Misconceptions about Colour and an Experimentally Tested
   Proposal to Combat them},
Journal = {International Journal of Science Education},
Year = {2013},
Volume = {35},
Number = {8},
Pages = {1299-1324},
Month = {JUN 1},
DOI = {10.1080/09500693.2013.770936},
ISSN = {0950-0693},
ResearcherID-Numbers = {Martinez Borreguero, Guadalupe/AAP-3295-2021
   Perez Rodriguez, Angel Luis/O-8039-2015
   Pardo, Pedro/D-3257-2011
   Suero Lopez, Maria Isabel/L-9237-2015
   },
ORCID-Numbers = {Martinez Borreguero, Guadalupe/0000-0002-0246-9406
   Perez Rodriguez, Angel Luis/0000-0001-8845-7424
   Pardo, Pedro/0000-0002-1603-9052
   Suero Lopez, Maria Isabel/0000-0003-1478-1795
   Sanchez-Fernandez, Gregorio Esteban/0000-0002-6059-7182},
Unique-ID = {WOS:000318278800002},
}

@article{ WOS:000292149900002,
Author = {Ebersbach, Mirjam and Van Dooren, Wim and Verschaffel, Lieven},
Title = {KNOWLEDGE ON ACCELERATED MOTION AS MEASURED BY IMPLICIT AND EXPLICIT
   TASKS IN 5 TO 16 YEAR OLDS},
Journal = {International Journal of Science and Mathematics Education},
Year = {2011},
Volume = {9},
Number = {1},
Pages = {25-46},
Month = {FEB},
DOI = {10.1007/s10763-010-9208-5},
ISSN = {1571-0068},
EISSN = {1573-1774},
ORCID-Numbers = {Van Dooren, Wim/0000-0001-5002-4340},
Unique-ID = {WOS:000292149900002},
}

@article{ WOS:000304504800049,
Author = {Ince, Elif},
Title = {The effectiveness of problem -based learning on students understanding
   of electromagnetic field and magnetism concepts},
Journal = {Energy Education Science and Technology Part B-Social and Educational
   Studies},
Year = {2012},
Volume = {4},
Number = {4},
Pages = {2383-2390},
Month = {OCT},
ISSN = {1308-7711},
ResearcherID-Numbers = {Ince, Elif/R-1127-2018},
Unique-ID = {WOS:000304504800049},
}

@article{ WOS:000373840100003,
Author = {Taslidere, Erdal},
Title = {Development and use of a three-tier diagnostic test to assess high
   school students' misconceptions about the photoelectric effect},
Journal = {Research in Science \& Technological Education},
Year = {2016},
Volume = {34},
Number = {2},
Pages = {164-186},
Month = {MAY 3},
DOI = {10.1080/02635143.2015.1124409},
ISSN = {0263-5143},
EISSN = {1470-1138},
ResearcherID-Numbers = {TASLIDERE, ERDAL/ABI-3943-2020},
Unique-ID = {WOS:000373840100003},
}

@article{ WOS:000309458400001,
Author = {Stewart, John and Miller, Mayo and Audo, Christine and Stewart, Gay},
Title = {Using cluster analysis to identify patterns in students' responses to
   contextually different conceptual problems},
Journal = {Physical Review Special Topics---Physics Education Research},
Year = {2012},
Volume = {8},
Number = {2},
Month = {OCT 4},
DOI = {10.1103/PhysRevSTPER.8.020112},
Article-Number = {020112},
ISSN = {1554-9178},
ORCID-Numbers = {Stewart, John/0000-0001-7325-5117},
Unique-ID = {WOS:000309458400001},
  eid={020112}
}

@article{ WOS:000297197200004,
Author = {Calik, Muammer and Okur, Murat and Taylor, Neil},
Title = {A Comparison of Different Conceptual Change Pedagogies Employed Within
   the Topic of ``Sound Propagation{''}},
Journal = {Journal of Science Education and Technology},
Year = {2011},
Volume = {20},
Number = {6},
Pages = {729-742},
Month = {DEC},
DOI = {10.1007/s10956-010-9266-z},
ISSN = {1059-0145},
EISSN = {1573-1839},
ResearcherID-Numbers = {CALIK, MUAMMER/AAA-1706-2019
   },
ORCID-Numbers = {Taylor, Neil/0000-0001-8438-319X
   CALIK, MUAMMER/0000-0001-8323-8783},
Unique-ID = {WOS:000297197200004},
}

@article{ WOS:001008829400001,
Author = {Aregehagn, Ehtegebreal and Lykknes, Annette and Getahun, Dawit Asrat and
   Febri, Maria I. M.},
Title = {Representation of Image Formation-Observation in Optics in {Ethiopian}
   Textbooks: Student Learning Difficulties as an Analytical Tool},
Journal = {Education Sciences},
Year = {2023},
Volume = {13},
Number = {5},
Month = {APR 26},
DOI = {10.3390/educsci13050445},
Article-Number = {445},
EISSN = {2227-7102},
ResearcherID-Numbers = {Lykknes, Annette/AAA-4129-2020
   Getahun, Dawit/GPC-4929-2022
   },
ORCID-Numbers = {Lykknes, Annette/0000-0002-1932-7754
   Getahun, Dawit Asrat/0000-0002-5733-0516},
Unique-ID = {WOS:001008829400001},
  eid={445}
}

@article{ WOS:000355004600008,
Author = {Ultay, Neslihan},
Title = {THE EFFECT OF CONCEPT CARTOONS EMBEDDED WITHIN CONTEXT-BASED CHEMISTRY: CHEMICAL BONDING},
Journal = {Journal of Baltic Science Education},
Year = {2015},
Volume = {14},
Number = {1},
Pages = {96-108},
ISSN = {1648-3898},
Unique-ID = {WOS:000355004600008},
DOI = {10.33225/jbse/15.14.96},
}

@article{ WOS:000485116000001,
Author = {Canlas, Ian Phil},
Title = {Using visual representations in identifying students' preconceptions in
   friction},
Journal = {Research in Science \& Technological Education},
Year = {2021},
Volume = {39},
Number = {2},
Pages = {156-184},
Month = {APR 3},
DOI = {10.1080/02635143.2019.1660630},
EarlyAccessDate = {SEP 2019},
ISSN = {0263-5143},
EISSN = {1470-1138},
ResearcherID-Numbers = {Canlas, Ian Phil/A-7412-2016},
ORCID-Numbers = {Canlas, Ian Phil/0000-0003-0056-3340},
Unique-ID = {WOS:000485116000001},
}

@article{ WOS:000469503400002,
Author = {Ferreira, Annalize and Lemmer, Miriam and Gunstone, Richard},
Title = {Alternative Conceptions: Turning Adversity into Advantage},
Journal = {Research in Science Education},
Year = {2019},
Volume = {49},
Number = {3},
Pages = {657-678},
Month = {JUN},
DOI = {10.1007/s11165-017-9638-y},
ISSN = {0157-244X},
EISSN = {1573-1898},
ResearcherID-Numbers = {Gunstone, Richard/ABE-3798-2020
   },
ORCID-Numbers = {Ferreira, Annalize/0000-0003-1223-5352
   Gunstone, Richard/0000-0002-0827-2663},
Unique-ID = {WOS:000469503400002},
}

@article{ WOS:000399912700003,
Author = {Nelson, Katherine G. and McKenna, Ann F. and Brem, Sarah K. and Hilpert,
   Jonathan and Husman, Jenefer and Pettinato, Eva},
Title = {Students' Misconceptions about Semiconductors and Use of Knowledge in
   Simulations},
Journal = {Journal of Engineering Education},
Year = {2017},
Volume = {106},
Number = {2},
Pages = {218-244},
Month = {APR},
DOI = {10.1002/jee.20163},
ISSN = {1069-4730},
EISSN = {2168-9830},
ResearcherID-Numbers = {McKenna, Ann/B-7602-2009
   husman, Jenefer/JXN-9607-2024
   },
ORCID-Numbers = {Hilpert, Jonathan/0000-0002-0173-1687},
Unique-ID = {WOS:000399912700003},
}

@article{ WOS:000741862900001,
Author = {Irmak, Meltem and Inaltun, Huseyin and Ercan-Dursun, Jale and
   Yanis-Kellec, Hilal and Yuruk, Nejla},
Title = {Development and Application of a Three-Tier Diagnostic Test to Assess
   Pre-service Science Teachers' Understanding on Work-Power and Energy
   Concepts},
Journal = {International Journal of Science and Mathematics Education},
Year = {2023},
Volume = {21},
Number = {1},
Pages = {159-185},
Month = {JAN},
DOI = {10.1007/s10763-021-10242-6},
EarlyAccessDate = {JAN 2022},
ISSN = {1571-0068},
EISSN = {1573-1774},
ResearcherID-Numbers = {Irmak, Meltem/AHD-5002-2022
   Yuruk, Nejla/AHD-7902-2022
   Yanış Kelleci, Hilal/AHD-5395-2022
   },
ORCID-Numbers = {INALTUN, HUSEYIN/0000-0002-1981-5539
   Yanis Kelleci, Hilal/0000-0002-9686-8538
   Yuruk, Nejla/0000-0001-9240-750X},
Unique-ID = {WOS:000741862900001},
}

@article{ WOS:000184915100005,
Author = {Jimoyiannis, A and Komis, V},
Title = {Investigating {Greek} students' ideas about forces and motion},
Journal = {Research in Science Education},
Year = {2003},
Volume = {33},
Number = {3},
Pages = {375-392},
DOI = {10.1023/A:1025457116654},
ISSN = {0157-244X},
ResearcherID-Numbers = {Jimoyiannis, Athanassios/AAD-7512-2021},
Unique-ID = {WOS:000184915100005},
}

@article{ WOS:000466445500001,
Author = {Eaton, Philip and Vavruska, Kinsey and Willoughby, Shannon},
Title = {Exploring the preinstruction and postinstruction non-{Newtonian} world
   views as measured by the {Force Concept Inventory}},
Journal = {Physical Review Physics Education Research},
Year = {2019},
Volume = {15},
Number = {1},
Month = {APR 25},
DOI = {10.1103/PhysRevPhysEducRes.15.010123},
Article-Number = {010123},
ISSN = {2469-9896},
ORCID-Numbers = {Willoughby, Shannon/0000-0002-5690-9489},
Unique-ID = {WOS:000466445500001},
  eid={010123}
}

@article{ WOS:A1996UH07600007,
Author = {Atwood, RK and Atwood, VA},
Title = {Preservice elementary teachers' conceptions of the causes of seasons},
Journal = {Journal of Research in Science Teaching},
Year = {1996},
Volume = {33},
Number = {5},
Pages = {553-563},
Month = {MAY},
ISSN = {0022-4308},
Unique-ID = {WOS:A1996UH07600007},
DOI = {10.1002/(SICI)1098-2736(199605)33:5<553::AID-TEA6>3.0.CO;2-Q},
}

@article{ WOS:001015331200003,
Author = {Afifah and Mufit, F. and Hamdi and Dewi, W. S.},
Title = {Interactive Multimedia Design Based on Cognitive Conflict on Dynamic
   Fluid Using {Adobe Animate CC} 2019},
Journal = {Jurnal Pendidikan Fisika Indonesia},
Year = {2023},
Volume = {19},
Number = {1},
Pages = {24-35},
Month = {JUN},
DOI = {10.15294/jpfi.v19i1.34777},
ISSN = {1693-1246},
EISSN = {2355-3812},
Unique-ID = {WOS:001015331200003},
}

@article{ WOS:000394440400004,
Author = {Atasoy, Sengul and Ergin, Serap},
Title = {The effect of concept cartoon-embedded worksheets on grade 9 students'
   conceptual understanding of {Newton}'s Laws of Motion},
Journal = {Research in Science \& Technological Education},
Year = {2017},
Volume = {35},
Number = {1},
Pages = {58-73},
Month = {MAR},
DOI = {10.1080/02635143.2016.1248926},
ISSN = {0263-5143},
EISSN = {1470-1138},
ResearcherID-Numbers = {Atasoy, Şengül/AGQ-2563-2022
   },
ORCID-Numbers = {Atasoy, Sengul/0000-0002-7664-1010},
Unique-ID = {WOS:000394440400004},
}

@article{ WOS:000574066100002,
Author = {Volfson, Alexander and Eshach, Haim and Ben-Abu, Yuval},
Title = {When Technology Meets Acoustics: Students' Ideas About the Underlying
   Principles Explaining Simple Acoustic Devices},
Journal = {Research in Science Education},
Year = {2021},
Volume = {51},
Number = {4},
Pages = {911-938},
Month = {AUG},
DOI = {10.1007/s11165-019-09913-w},
EarlyAccessDate = {JAN 2020},
ISSN = {0157-244X},
EISSN = {1573-1898},
ResearcherID-Numbers = {Abu, Yuval/AAD-1402-2020},
Unique-ID = {WOS:000574066100002},
}

@article{ WOS:000253202500001,
Author = {Hsu, Ying-Shao and Wu, Hsin-Kai and Hwang, Fu-Kwun},
Title = {Fostering high school students' conceptual understandings about seasons:
   The design of a technology-enhanced learning environment},
Journal = {Research in Science Education},
Year = {2008},
Volume = {38},
Number = {2},
Pages = {127-147},
Month = {MAR},
DOI = {10.1007/s11165-007-9041-1},
ISSN = {0157-244X},
EISSN = {1573-1898},
ResearcherID-Numbers = {Wu, Hsin-Kai/B-7260-2015},
ORCID-Numbers = {Hsu, Ying-Shao/0000-0002-1635-8213
   Wu, Hsin-Kai/0000-0003-0018-9969},
Unique-ID = {WOS:000253202500001},
}

@article{ WOS:000179612600005,
Author = {Eryilmaz, A},
Title = {Effects of conceptual assignments and conceptual change discussions on
   students' misconceptions and achievement regarding force and motion},
Journal = {Journal of Research in Science Teaching},
Year = {2002},
Volume = {39},
Number = {10},
Pages = {1001-1015},
Month = {DEC},
DOI = {10.1002/tea.10054},
ISSN = {0022-4308},
EISSN = {1098-2736},
ResearcherID-Numbers = {ERYILMAZ, ALI/C-3398-2013},
ORCID-Numbers = {ERYILMAZ, ALI/0000-0003-2161-6018},
Unique-ID = {WOS:000179612600005},
}

@article{ WOS:000322006100004,
Author = {Dega, Bekele Gashe and Kriek, Jeanne and Mogese, Temesgen Fereja},
Title = {Students' conceptual change in electricity and magnetism using
   simulations: A comparison of cognitive perturbation and cognitive
   conflict},
Journal = {Journal of Research in Science Teaching},
Year = {2013},
Volume = {50},
Number = {6, SI},
Pages = {677-698},
Month = {AUG},
DOI = {10.1002/tea.21096},
ISSN = {0022-4308},
EISSN = {1098-2736},
ResearcherID-Numbers = {Kriek, Jeanne/E-1312-2015
   },
ORCID-Numbers = {Kriek, Jeanne/0000-0001-6248-4563
   Dega, Bekele Gashe/0000-0002-2115-1348},
Unique-ID = {WOS:000322006100004},
}

@article{ WOS:000305848300002,
Author = {Prince, Michael and Vigeant, Margot and Nottis, Katharyn},
Title = {Development of the {Heat and Energy Concept Inventory}: Preliminary
   Results on the Prevalence and Persistence of Engineering Students'
   Misconceptions},
Journal = {Journal of Engineering Education},
Year = {2012},
Volume = {101},
Number = {3},
Pages = {412-438},
Month = {JUL},
DOI = {10.1002/j.2168-9830.2012.tb00056.x},
ISSN = {1069-4730},
ORCID-Numbers = {Vigeant, Margot/0000-0003-2593-7097},
Unique-ID = {WOS:000305848300002},
}

@article{ WOS:001040148700001,
Author = {Bhaw, N. and Kriek, J. and Lemmer, M.},
Title = {Insights from coherence in students' scientific reasoning skills},
Journal = {Heliyon},
Year = {2023},
Volume = {9},
Number = {7},
Month = {JUL},
DOI = {10.1016/j.heliyon.2023.e17349},
EarlyAccessDate = {JUL 2023},
Article-Number = {e17349},
EISSN = {2405-8440},
Unique-ID = {WOS:001040148700001},
  eid={e17349}
}

@article{ WOS:000749487700001,
Author = {Wang, Qian and Zhu, Yanmei and Wei, Lili and Deng, Huihua},
Title = {Visual Attention Pattern of Middle School Students During
   Problem-Solving in Physics},
Journal = {Mind Brain and Education},
Year = {2022},
Volume = {16},
Number = {2, SI},
Pages = {99-111},
Month = {MAY},
DOI = {10.1111/mbe.12313},
EarlyAccessDate = {FEB 2022},
ISSN = {1751-2271},
EISSN = {1751-228X},
Unique-ID = {WOS:000749487700001},
}

@article{ WOS:000269975100007,
Author = {Gonen, Selahattin},
Title = {A Study on Student Teachers' Misconceptions and Scientifically
   Acceptable Conceptions About Mass and Gravity},
Journal = {Journal of Science Education and Technology},
Year = {2008},
Volume = {17},
Number = {1},
Pages = {70-81},
Month = {FEB},
DOI = {10.1007/s10956-007-9083-1},
ISSN = {1059-0145},
EISSN = {1573-1839},
Unique-ID = {WOS:000269975100007},
}

@article{ WOS:000705678200001,
Author = {Taban, Tugce and Kiray, Seyit Ahmet},
Title = {Determination of Science Teacher Candidates' Misconceptions on Liquid
   Pressure with Four-Tier Diagnostic Test},
Journal = {International Journal of Science and Mathematics Education},
Year = {2022},
Volume = {20},
Number = {8},
Pages = {1791-1811},
Month = {DEC},
DOI = {10.1007/s10763-021-10224-8},
EarlyAccessDate = {OCT 2021},
ISSN = {1571-0068},
EISSN = {1573-1774},
ResearcherID-Numbers = {Kiray, Seyit Ahmet/AGQ-4408-2022},
Unique-ID = {WOS:000705678200001},
}

@article{ WOS:000207961100003,
Author = {Bayraktar, Sule},
Title = {MISCONCEPTIONS OF {Turkish} PRE-SERVICE TEACHERS ABOUT FORCE AND MOTION},
Journal = {International Journal of Science and Mathematics Education},
Year = {2009},
Volume = {7},
Number = {2},
Pages = {273-291},
Month = {APR},
DOI = {10.1007/s10763-007-9120-9},
ISSN = {1571-0068},
EISSN = {1573-1774},
ResearcherID-Numbers = {Bayraktar, Sule/ABB-1295-2021},
Unique-ID = {WOS:000207961100003},
}

@article{ WOS:000766923500011,
Author = {Wang, Jue and Liang, Kaihua},
Title = {A Cognitive Diagnosis Method in Adaptive Learning System Based on
   Preconceptions},
Journal = {Scientific Programming},
Year = {2022},
Volume = {2022},
Month = {JAN 10},
DOI = {10.1155/2022/5011804},
Article-Number = {5011804},
ISSN = {1058-9244},
EISSN = {1875-919X},
ORCID-Numbers = {Liang, Kaihua/0000-0002-8813-194X
   WANG, Jue/0000-0002-3004-0877},
Unique-ID = {WOS:000766923500011},
  eid={5011804}
}

@article{ WOS:000423523700001,
Author = {Scott, Terry F. and Schumayer, Daniel},
Title = {Central distractors in {Force Concept Inventory} data},
Journal = {Physical Review Physics Education Research},
Year = {2018},
Volume = {14},
Number = {1},
Month = {JAN 30},
DOI = {10.1103/PhysRevPhysEducRes.14.010106},
Article-Number = {010106},
ISSN = {2469-9896},
ORCID-Numbers = {Schumayer, Daniel/0000-0003-4797-1712},
Unique-ID = {WOS:000423523700001},
  eid={010106}
}

@article{ WOS:000184928700003,
Author = {Palmer, DH},
Title = {Investigating the relationship between refutational text and conceptual
   change},
Journal = {Science Education},
Year = {2003},
Volume = {87},
Number = {5},
Pages = {663-684},
Month = {SEP},
DOI = {10.1002/sce.1056},
ISSN = {0036-8326},
Unique-ID = {WOS:000184928700003},
}

@article{ WOS:000428300400001,
Author = {Adadan, Emine and Yavuzkaya, Merve Nur},
Title = {Examining the progression and consistency of thermal concepts: a
   cross-age study},
Journal = {International Journal of Science Education},
Year = {2018},
Volume = {40},
Number = {4},
Pages = {371-396},
DOI = {10.1080/09500693.2018.1423711},
ISSN = {0950-0693},
EISSN = {1464-5289},
ResearcherID-Numbers = {Adadan, Emine/F-9328-2011
   },
ORCID-Numbers = {Adadan, Emine/0000-0002-7109-4935
   Yavuzkaya, Merve/0000-0003-4230-2026},
Unique-ID = {WOS:000428300400001},
}

@article{ WOS:000176433100001,
Author = {Winer, Gerald A. and Cottrell, Jane E. and Gregg, Virginia and Fournier, Jody S. and Bica, Lori A.},
Title = {Fundamentally misunderstanding visual perception - Adults' belief in
   visual emissions},
Journal = {American Psychologist},
Year = {2002},
Volume = {57},
Number = {6-7},
Pages = {417-424},
Month = {JUN-JUL},
DOI = {10.1037//0003-066X.57.6-7.417},
ISSN = {0003-066X},
EISSN = {1935-990X},
Unique-ID = {WOS:000176433100001},
}

@article{ WOS:000088224100002,
Author = {Clerk, D and Rutherford, M},
Title = {Language as a confounding variable in the diagnosis of misconceptions},
Journal = {International Journal of Science Education},
Year = {2000},
Volume = {22},
Number = {7},
Pages = {703-717},
Month = {JUL},
DOI = {10.1080/09500690050044053},
ISSN = {0950-0693},
Unique-ID = {WOS:000088224100002},
}

@article{ WOS:000287470500024,
Author = {Atasoy, Sengul and Kucuk, Mehmet and Akdeniz, Ali Riza},
Title = {Remedying science student teachers' misconceptions of force and motion
   using worksheets based on constructivist learning theory},
Journal = {Energy Education Science and Technology Part B-Social and Educational
   Studies},
Year = {2011},
Volume = {3},
Number = {4},
Pages = {653-668},
Month = {OCT},
ISSN = {1308-7711},
ResearcherID-Numbers = {Atasoy, Şengül/AGQ-2563-2022
   KÜÇÜK, Mehmet/HNB-5023-2023
   },
ORCID-Numbers = {KUCUK, MEHMET/0000-0001-5910-4099
   Atasoy, Sengul/0000-0002-7664-1010},
Unique-ID = {WOS:000287470500024},
}

@article{ WOS:000970595800025,
Author = {Wancham, Kittitas and Tangdhanakanond, Kamonwan and Kanjanawasee,
   Sirichai},
Title = {Sex and Grade Issues in Influencing Misconceptions about Force and Laws
   of Motion: An Application of Cognitively Diagnostic Assessment},
Journal = {International Journal of Instruction},
Year = {2023},
Volume = {16},
Number = {2},
Pages = {437-456},
Month = {APR},
DOI = {10.29333/iji.2023.16224a},
ISSN = {1694-609X},
EISSN = {1308-1470},
ORCID-Numbers = {Wancham, Kittitas/0000-0002-9930-3428},
Unique-ID = {WOS:000970595800025},
}

@article{ WOS:000885770300001,
Author = {Wheatley, Christopher and Wells, James and Pritchard, David E. and
   Stewart, John},
Title = {Comparing conceptual understanding across institutions with module
   analysis},
Journal = {Physical Review Physics Education Research},
Year = {2022},
Volume = {18},
Number = {2},
Month = {NOV 14},
DOI = {10.1103/PhysRevPhysEducRes.18.020132},
Article-Number = {020132},
ISSN = {2469-9896},
ORCID-Numbers = {Stewart, John/0000-0001-7325-5117},
Unique-ID = {WOS:000885770300001},
  eid={020132}
}

@article{ WOS:000367386000001,
Author = {Aviani, Ivica and Erceg, Natasa and Mesic, Vanes},
Title = {Drawing and using free body diagrams: Why it may be better not to
   decompose forces},
Journal = {Physical Review Special Topics---Physics Education Research},
Year = {2015},
Volume = {11},
Number = {2},
Month = {DEC 29},
DOI = {10.1103/PhysRevSTPER.11.020137},
Article-Number = {020137},
ISSN = {1554-9178},
ResearcherID-Numbers = {Aviani, Ivica/F-5059-2017
   Mesic, Vanes/M-6766-2019
   Erceg, Natasa/O-7573-2018
   },
ORCID-Numbers = {Erceg, Natasa/0000-0003-3308-1372
   Aviani, Ivica/0000-0002-5002-388X},
Unique-ID = {WOS:000367386000001},
  eid={020137}
}

@article{ WOS:000608678900001,
Author = {Wheatley, Christopher and Wells, James and Henderson, Rachel and
   Stewart, John},
Title = {Applying module analysis to the Conceptual Survey of Electricity and
   Magnetism},
Journal = {Physical Review Physics Education Research},
Year = {2021},
Volume = {17},
Number = {1},
Month = {JAN 19},
DOI = {10.1103/PhysRevPhysEducRes.17.010102},
Article-Number = {010102},
ISSN = {2469-9896},
ORCID-Numbers = {Wells, James/0000-0002-8304-8258
   Stewart, John/0000-0001-7325-5117},
Unique-ID = {WOS:000608678900001},
  eid={010102}
}

@incollection{ WOS:000568612900004,
Author = {Neidorf, Teresa and Arora, Alka and Erberber, Ebru and Tsokodayi,
   Yemurai and Mai, Thanh},
Book-Author = {Neidorf, T
   Arora, A
   Erberber, E
   Tsokodayi, Y
   Mai, T},
Title = {Methodology Used to Analyze Student Misconceptions, Errors, and
   Misunderstandings in {TIMSS}},
Booktitle = {Student Misconceptions and Errors in Physics and Mathematics: Exploring
   Data from TIMSS and TIMSS Advanced},
Series = {IEA Research for Education},
Year = {2020},
Volume = {9},
Pages = {21-35},
DOI = {10.1007/978-3-030-30188-0_3},
ISSN = {2366-1631},
EISSN = {2366-164X},
ISBN = {978-3-030-30188-0; 978-3-030-30187-3},
Unique-ID = {WOS:000568612900004},
Publisher = {Springer},
Address = {Cham},
Chapter = {3},
}

@article{ WOS:000281414000007,
Author = {Salazar, Agustin and Apinaniz, Estibaliz and Mendioroz, Arantza and
   Oleaga, Alberto},
Title = {A thermal paradox: which gets warmer?},
Journal = {European Journal of Physics},
Year = {2010},
Volume = {31},
Number = {5},
Pages = {1053-1059},
Month = {SEP},
DOI = {10.1088/0143-0807/31/5/007},
ISSN = {0143-0807},
ResearcherID-Numbers = {Mendioroz, Arantza/D-3701-2015
   Apinaniz, Estibaliz/D-8265-2015
   Salazar, Agustin/K-5814-2014
   Oleaga, Alberto/L-2873-2014
   },
ORCID-Numbers = {Apinaniz, Estibaliz/0000-0002-3668-8316
   Salazar, Agustin/0000-0002-4108-332X
   Oleaga, Alberto/0000-0002-3618-9103
   MENDIOROZ ASTIGARRAGA, MARIA ARANZAZU/0000-0002-8421-8896},
Unique-ID = {WOS:000281414000007},
}

@article{ WOS:000329795900010,
Author = {Demirezen, Selcuk and Yagbasan, Rahmi},
Title = {The Effect of {7E} Model on Misconceptions About Simple Electrical
   Circuits},
Journal = {Hacettepe University Journal of Education},
Year = {2013},
Volume = {28},
Number = {2},
Pages = {132-151},
ISSN = {2536-4758},
ResearcherID-Numbers = {demirezen, selçuk/ABC-1756-2021},
Unique-ID = {WOS:000329795900010},
}

@article{ WOS:000742147000003,
Author = {Kizilcik, Hasan Sahin and Aygun, Muge and Sahin, Esin and
   Onder-Celikkanli, Nuray and Turk, Osman and Taskin, Tugba and Gunes,
   Bilal},
Title = {Possible misconceptions about solid friction},
Journal = {Physical Review Physics Education Research},
Year = {2021},
Volume = {17},
Number = {2},
Month = {NOV 12},
DOI = {10.1103/PhysRevPhysEducRes.17.023107},
Article-Number = {023107},
ISSN = {2469-9896},
ResearcherID-Numbers = {Taşkın, Tuğba/AHB-5750-2022
   Onder Celikkanli, Nuray/ABR-6907-2022
   Aygun, Muge/JJF-7781-2023
   GUNES, Prof. Dr. Bilal/A-4520-2017
   Kizilcik, Hasan Sahin/A-7692-2018
   },
ORCID-Numbers = {ONDER CELIKKANLI, NURAY/0000-0002-2726-4386
   Aygun, Muge/0000-0002-5268-2205
   Taskin, Tugba/0000-0002-8738-0012
   Turk, Osman/0000-0001-9429-406X
   GUNES, Prof. Dr. Bilal/0000-0002-3911-6778
   Kizilcik, Hasan Sahin/0000-0001-8622-0765
   SAHIN, Esin/0000-0001-6506-1507},
Unique-ID = {WOS:000742147000003},
  eid={023107}
}

@article{ WOS:000285629700012,
Author = {Yilmaz, Serkan},
Title = {Investigating Prospective Elementary Teachers' Understanding of
   Analogical Cases in {Newton}'s Third Law},
Journal = {Eurasian Journal of Educational Research},
Year = {2010},
Volume = {10},
Number = {41},
Pages = {201-219},
Month = {FAL},
ISSN = {1302-597X},
EISSN = {2528-8911},
ResearcherID-Numbers = {YILMAZ, SERKAN/I-8770-2013},
ORCID-Numbers = {YILMAZ, SERKAN/0000-0003-1800-0765},
Unique-ID = {WOS:000285629700012},
}

@article{ WOS:000405956500015,
Author = {Cibik, Ayse Sert},
Title = {Determining Science Teacher Candidates' Academic Knowledge and
   Misconceptions about Electric Current},
Journal = {Educational Sciences: Theory and Practice},
Year = {2017},
Volume = {17},
Number = {3},
Pages = {1061-1090},
Month = {JUN},
DOI = {10.12738/estp.2017.3.0530},
ISSN = {1303-0485},
EISSN = {2148-7561},
Unique-ID = {WOS:000405956500015},
}

@article{ WOS:000355686800001,
Author = {Passelaigue, Dominique and Munier, Valerie},
Title = {Schoolteacher trainees' difficulties about the concepts of attribute and
   measurement},
Journal = {Educational Studies in Mathematics},
Year = {2015},
Volume = {89},
Number = {3},
Pages = {307-336},
Month = {JUL},
DOI = {10.1007/s10649-015-9610-6},
ISSN = {0013-1954},
EISSN = {1573-0816},
Unique-ID = {WOS:000355686800001},
}

@article{ WOS:000354532600005,
Author = {Gurcay, Deniz and Gulbas, Etna},
Title = {Development of three-tier heat, temperature and internal energy
   diagnostic test},
Journal = {Research in Science \& Technological Education},
Year = {2015},
Volume = {33},
Number = {2},
Pages = {197-217},
Month = {MAY 4},
DOI = {10.1080/02635143.2015.1018154},
ISSN = {0263-5143},
EISSN = {1470-1138},
Unique-ID = {WOS:000354532600005},
}

@article{ WOS:000751558100012,
Author = {Kaniawati, Ida and Maulidina, Widyatami Nurul and Novia, Hera and Suyana, Iyon and Samsudin, Achmad and Aminudin, Adam Hadiana and Suhendi, Endi},
Title = {Implementation of Interactive Conceptual Instruction ({ICI}) Learning
   Model Assisted by Computer Simulation: Impact of Students' Conceptual
   Changes on Force and Vibration},
Journal = {International Journal of Emerging Technologies in Learning},
Year = {2021},
Volume = {16},
Number = {22},
Pages = {167-188},
DOI = {10.3991/ijet.v16i22.25465},
ISSN = {1863-0383},
ResearcherID-Numbers = {Aminudin, Adam Hadiana/S-7982-2018
   Samsudin, Achmad/E-5170-2015
   Suhendi, Endi/E-3390-2016},
ORCID-Numbers = {Hadiana Aminudin, Adam/0000-0001-7409-9195
   Samsudin, Achmad/0000-0003-3564-6031
   Suhendi, Endi/0000-0001-8256-8537},
Unique-ID = {WOS:000751558100012},
}

@article{ WOS:001163538600002,
Author = {Yasuda, Jun-ichiro and Hull, Michael M. and Mae, Naohiro},
Title = {Visualizing depth of student conceptual understanding using subquestions
   and alluvial diagrams},
Journal = {Physical Review Physics Education Research},
Year = {2023},
Volume = {19},
Number = {2},
Month = {AUG 31},
DOI = {10.1103/PhysRevPhysEducRes.19.020121},
Article-Number = {020121},
ISSN = {2469-9896},
ORCID-Numbers = {Mae, Naohiro/0000-0001-5118-299X
   Hull, Michael M./0000-0003-1551-9553},
Unique-ID = {WOS:001163538600002},
  eid={020121}
}

@article{ WOS:000442716200001,
Author = {Donovan, Amalia M. and Zhan, Jennifer and Rapp, David N.},
Title = {Supporting historical understandings with refutation texts},
Journal = {Contemporary Educational Psychology},
Year = {2018},
Volume = {54},
Pages = {1-11},
Month = {JUL},
DOI = {10.1016/j.cedpsych.2018.04.002},
ISSN = {0361-476X},
EISSN = {1090-2384},
Unique-ID = {WOS:000442716200001},
}

@article{ WOS:000076967200002,
Author = {Roschelle, J},
Title = {Beyond romantic versus sceptic: a microanalysis of conceptual change in
   kinematics},
Journal = {International Journal of Science Education},
Year = {1998},
Volume = {20},
Number = {9},
Pages = {1025-1042},
Month = {NOV},
DOI = {10.1080/0950069980200902},
ISSN = {0950-0693},
ResearcherID-Numbers = {Roschelle, Jeremy/AAJ-6292-2020
   },
ORCID-Numbers = {Roschelle, Jeremy/0000-0003-2219-0506},
Unique-ID = {WOS:000076967200002},
}

@article{ WOS:000318658000001,
Author = {Aslanides, J. S. and Savage, C. M.},
Title = {Relativity concept inventory: Development, analysis, and results},
Journal = {Physical Review Special Topics---Physics Education Research},
Year = {2013},
Volume = {9},
Number = {1},
Month = {MAY 9},
DOI = {10.1103/PhysRevSTPER.9.010118},
Article-Number = {010118},
ISSN = {1554-9178},
ResearcherID-Numbers = {Savage, Craig/A-1424-2008
   },
ORCID-Numbers = {Savage, Craig/0000-0003-1768-4982},
Unique-ID = {WOS:000318658000001},
  eid={010118}
}

@article{ WOS:000279714800003,
Author = {Yilmaz, Serkan and Eryilmaz, Ali},
Title = {Integrating Gender and Group Differences into Bridging Strategy},
Journal = {Journal of Science Education and Technology},
Year = {2010},
Volume = {19},
Number = {4},
Pages = {341-355},
Month = {AUG},
DOI = {10.1007/s10956-010-9204-0},
ISSN = {1059-0145},
ResearcherID-Numbers = {YILMAZ, SERKAN/I-8770-2013
   ERYILMAZ, ALI/C-3398-2013},
ORCID-Numbers = {YILMAZ, SERKAN/0000-0003-1800-0765
   ERYILMAZ, ALI/0000-0003-2161-6018},
Unique-ID = {WOS:000279714800003},
}

@article{ WOS:000311506200004,
Author = {Pejuan, Arcadi and Bohigas, Xavier and Jaen, Xavier and Periago,
   Cristina},
Title = {Misconceptions About Sound Among Engineering Students},
Journal = {Journal of Science Education and Technology},
Year = {2012},
Volume = {21},
Number = {6},
Pages = {669-685},
Month = {DEC},
DOI = {10.1007/s10956-011-9356-6},
ISSN = {1059-0145},
EISSN = {1573-1839},
ResearcherID-Numbers = {Bohigas, Xavier/JVD-9361-2023
   Periago, Cristina/H-8996-2015
   },
ORCID-Numbers = {Periago, Cristina/0000-0002-4858-7926
   Jaen, Xavier/0000-0002-0676-3848},
Unique-ID = {WOS:000311506200004},
}

@article{ WOS:000228024600001,
Author = {Chi, M. T. H.},
Title = {Commonsense conceptions of emergent processes: Why some misconceptions
   are robust},
Journal = {Journal of the Learning Sciences},
Year = {2005},
Volume = {14},
Number = {2},
Pages = {161-199},
DOI = {10.1207/s15327809jls1402\_1},
ISSN = {1050-8406},
EISSN = {1532-7809},
Unique-ID = {WOS:000228024600001},
}

@article{ WOS:000297176100001,
Author = {Rosenblatt, Rebecca and Heckler, Andrew F.},
Title = {Systematic study of student understanding of the relationships between
   the directions of force, velocity, and acceleration in one dimension},
Journal = {Physical Review Special Topics---Physics Education Research},
Year = {2011},
Volume = {7},
Number = {2},
Month = {NOV 2},
DOI = {10.1103/PhysRevSTPER.7.020112},
Article-Number = {020112},
ISSN = {1554-9178},
ResearcherID-Numbers = {Heckler, Andrew/A-7291-2010},
Unique-ID = {WOS:000297176100001},
  eid={020112}
}

@article{ WOS:000357865000001,
Author = {Lee, Hyunju and Schneider, Stephen E.},
Title = {Using astronomical photographs to investigate misconceptions about
   galaxies and spectra: Question development for clicker use},
Journal = {Physical Review Special Topics---Physics Education Research},
Year = {2015},
Volume = {11},
Number = {2},
Month = {JUL 13},
DOI = {10.1103/PhysRevSTPER.11.020101},
Article-Number = {020101},
ISSN = {1554-9178},
ResearcherID-Numbers = {Lee, Hyunju/S-6130-2019
   },
ORCID-Numbers = {Lee, Hyunju/0000-0001-9408-2630},
Unique-ID = {WOS:000357865000001},
  eid={020101}
}

@article{ WOS:000600006900001,
Author = {Kaliampos, George and Ravanis, Konstantinos and Vavougios, Denis},
Title = {A comparison study of alternative conceptions on impetus theory and
   projectile motion of adolescents with typical development and high
   functioning autism spectrum disorder},
Journal = {International Journal of Science Education},
Year = {2021},
Volume = {43},
Number = {1},
Pages = {128-156},
Month = {JAN 2},
DOI = {10.1080/09500693.2020.1859156},
EarlyAccessDate = {DEC 2020},
ISSN = {0950-0693},
EISSN = {1464-5289},
ResearcherID-Numbers = {Kaliampos, George/MFH-5575-2025
   Ravanis, Konstantinos/V-1267-2018},
Unique-ID = {WOS:000600006900001},
}

@article{docktor2014synthesis,
  title={Synthesis of discipline-based education research in physics},
  author={Docktor, Jennifer L and Mestre, Jos{\'e} P},
  journal={Physical Review Special Topics---Physics Education Research},
  volume={10},
  number={2},
  pages={020119},
  year={2014},
  publisher={APS}
}

@article{siddiqui2017diverse,
  author={Siddiqui, Shabnam and Singh, Chandralekha},
  title={How diverse are physics instructors' attitudes and approaches to teaching undergraduate level quantum mechanics?},
  journal={European Journal of Physics},
  volume={38}, number={3}, pages={035703}, year={2017},
  doi={10.1088/1361-6404/aa6131}
}

@article{meyer2022interdisciplinary,
  author={Meyer, Josephine C. and Passante, Gina and Pollock, Steven J. and Wilcox, Bethany R.},
  title={Today's interdisciplinary quantum information classroom: Themes from a survey of quantum information science instructors},
  journal={Physical Review Physics Education Research},
  volume={18}, number={1}, pages={010150}, year={2022},
  doi={10.1103/PhysRevPhysEducRes.18.010150}
}

@article{meyer2023media,
  author={Meyer, Josephine C. and Pollock, Steven James and Wilcox, Bethany R. and Passante, Gina},
  title={How media hype affects our physics teaching: A case study on quantum computing},
  journal={The Physics Teacher},
  volume={61}, number={5}, pages={339--342}, year={2023},
  doi={10.1119/5.0117671}
}

@article{kruse2025characterization,
  author={Kruse, Jesse and Griston, Molly and Wilcox, Bethany R.},
  title={Characterization of upper-level undergraduate quantum mechanics courses in the {U.S.}},
  journal={Physical Review Physics Education Research},
  volume={21}, number={2}, pages={020162}, year={2025},
  doi={10.1103/63s3-7tb8}
}

@misc{waterloo2023undergraduate,
  author={{University of Waterloo}},
  title={{Undergraduate Studies Academic Calendar 2023--2024}},
  year={2023},
  url={https://academic-calendar-archive.uwaterloo.ca/undergraduate-studies/2023-2024/page/uWaterloo-Undergraduate-Calendar-Access.html},
  note={Archived academic calendar; accessed September 2026}
}

@misc{waterloo2023graduate,
  author={{University of Waterloo}},
  title={{Graduate Studies Academic Calendar, Fall 2023}},
  year={2023},
  url={https://academic-calendar-archive.uwaterloo.ca/graduate-studies/archive-fall-2023/index.html},
  note={Archived academic calendar; accessed September 2026}
}

@article{majidy2024noncommuting,
  title={Noncommuting charges can remove non-stationary quantum many-body dynamics},
  author={Majidy, Shayan},
  journal={Nature Communications},
  volume={15},
  number={1},
  pages={8246},
  year={2024},
  publisher={Nature Publishing Group UK London}
}

@article{majidy2019exploration,
  title={Exploration of an augmented set of Leggett-Garg inequalities using a noninvasive continuous-in-time velocity measurement},
  author={Majidy, Shayan-Shawn and Katiyar, Hemant and Anikeeva, Galit and Halliwell, Jonathan and Laflamme, Raymond},
  journal={Physical Review A},
  volume={100},
  number={4},
  pages={042325},
  year={2019},
  publisher={APS}
}

@article{majidy2021detecting,
  title={Detecting violations of macrorealism when the original Leggett-Garg inequalities are satisfied},
  author={Majidy, Shayan and Halliwell, Jonathan J and Laflamme, Raymond},
  journal={Physical Review A},
  volume={103},
  number={6},
  pages={062212},
  year={2021},
  publisher={APS}
}

@article{majidy2023unification,
  title={A Unification of the Coding Theory and OAQEC Perspectives on Hybrid Codes},
  author={Majidy, Shayan},
  journal={International Journal of Theoretical Physics},
  volume={62},
  number={8},
  pages={177},
  year={2023},
  publisher={Springer}
}

@book{majidy2024building,
  title={Building Quantum Computers: A Practical Introduction},
  author={Majidy, Shayan and Wilson, Christopher and Laflamme, Raymond},
  year={2024},
  publisher={Cambridge University Press}
}

@article{yunger2022build,
  title={How to build Hamiltonians that transport noncommuting charges in quantum thermodynamics},
  author={Yunger Halpern, Nicole and Majidy, Shayan},
  journal={npj Quantum Information},
  volume={8},
  number={1},
  pages={10},
  year={2022},
  publisher={Nature Publishing Group UK London}
}

@article{majidy2023non,
  title={Non-Abelian symmetry can increase entanglement entropy},
  author={Majidy, Shayan and Lasek, Aleksander and Huse, David A and Halpern, Nicole Yunger},
  journal={Physical Review B},
  volume={107},
  number={4},
  pages={045102},
  year={2023},
  publisher={APS}
}

@article{majidy2023critical,
  title={Critical phase and spin sharpening in SU (2)-symmetric monitored quantum circuits},
  author={Majidy, Shayan and Agrawal, Utkarsh and Gopalakrishnan, Sarang and Potter, Andrew C and Vasseur, Romain and Halpern, Nicole Yunger},
  journal={Physical Review B},
  volume={108},
  number={5},
  pages={054307},
  year={2023},
  publisher={APS}
}

@article{majidy2023noncommuting,
  title={Noncommuting conserved charges in quantum thermodynamics and beyond},
  author={Majidy, Shayan and Braasch Jr, William F and Lasek, Aleksander and Upadhyaya, Twesh and Kalev, Amir and Yunger Halpern, Nicole},
  journal={Nature Reviews Physics},
  volume={5},
  number={11},
  pages={689--698},
  year={2023},
  publisher={Nature Publishing Group UK London}
}

@mastersthesis{majidy2019violation,
  title={Violation of an augmented set of Leggett-Garg inequalities and the implementation of a continuous in time velocity measurement},
  author={Majidy, Shayan-Shawn},
  year={2019},
  school={University of Waterloo}
}
\interlinepenalty=\savedBibliographyInterlinePenalty\relax
\let\hyper@linkurl\bibliography@original@linkurl
\makeatother
\bibliographystyle{apsrev4-2-local}

\section*{Declarations}

\noindent\textbf{Author contributions.} SM conducted the study, performed the substantive analysis, and wrote the manuscript. Advisory input is described in Section~\ref{sec:interviewanalysis}. The interviews were transcribed with TurboScribe and checked by SM.

\noindent\textbf{Funding.} No project-specific funding was received. General doctoral support from the Vanier CGS is acknowledged above. 

\noindent\textbf{Data availability.} The author retains the checked interview transcripts privately. Complete transcripts and identifying links cannot be made public under the confidentiality agreements. Audio recordings were deleted after transcript checking and resolution of the remaining transcription uncertainties. The final recording was deleted in March 2025. The literature database and summaries are provided in Appendix~\ref{add:database}.

\noindent\textbf{Ethics approval and consent to participate.} The study involved voluntary interviews with adult university physics educators about their professional teaching experiences. All participants provided informed consent before taking part. The research followed the University of Waterloo's guidelines for ethical research, and its Office of Research Ethics confirmed that formal ethics approval was not required for this type of educational study. The consent and confidentiality arrangements described above remained applicable regardless of that determination.

\noindent\textbf{Consent to publish.} All twelve participants permitted publication of de-identified excerpts and coded paraphrases. The eleven instructors named in the acknowledgments explicitly agreed to acknowledgment by name; the twelfth requested anonymity. Acknowledgment of participation does not imply endorsement of the report or its interpretations.

\noindent\textbf{Competing interests.} The author declares no competing interests.

\noindent\textbf{Revision note.} This version substantially revises arXiv:2405.20923v2 (2025) and reports the same twelve-instructor interview project; no new interviews were conducted. The methods section now reports additional details of the original recruitment, interviewing, transcript checking, and analysis, and the ethics, consent, and data-handling arrangements are described more explicitly. The earlier description ``professionally transcribed'' is corrected to specify TurboScribe transcription followed by author checking against the recordings. References, literature summaries, category assignments, and scientific explanations have been corrected or revised. The literature component is now presented as a selected overview of 122 distinct works; the exclusions and counting corrections are explained in Section~\ref{sec:search}. Course-count changes reflect title-based regrouping, correction of the earlier counting, and removal of the erroneously included ECE 457C, not changes in institutional offerings. The instructional-planning table is a new author synthesis, not additional interview findings. The earlier classical--quantum comparative inference is not retained, as explained above. AI assistance during revision is described in the Acknowledgments.

\clearpage
\onecolumngrid
\appendix
\begingroup
\small
\hypersetup{citecolor=black,linkcolor=black}
\setlength{\tabcolsep}{6pt}
\renewcommand{\arraystretch}{1.12}
\setlength{\LTleft}{0pt}
\setlength{\LTright}{0pt}
\setlength{\LTpre}{7pt}
\setlength{\LTpost}{12pt}
\setlength{\LTcapwidth}{\textwidth}
\setlength{\extrarowheight}{0pt}
\newlength{\AppendixRefWidth}
\setlength{\AppendixRefWidth}{0.07\textwidth}
\newlength{\AppendixSummaryWidth}
\setlength{\AppendixSummaryWidth}{\dimexpr\textwidth-\AppendixRefWidth-2\tabcolsep\relax}
\newlength{\AppendixCodeWidth}
\setlength{\AppendixCodeWidth}{0.30\textwidth}
\newlength{\AppendixCourseWidth}
\setlength{\AppendixCourseWidth}{\dimexpr\textwidth-\AppendixCodeWidth-2\tabcolsep\relax}
\newcommand{\AppendixGroup}[1]{%
  \multicolumn{2}{@{}l@{}}{\rule{0pt}{18pt}\textbf{#1}}\\*[3pt]}
\newcommand{\AppendixTableNote}[1]{%
  \noindent\begin{minipage}{\textwidth}\small\RaggedRight #1\end{minipage}\par\medskip}

\makeatletter
\long\def\LT@makecaption#1#2#3{%
  \LT@mcol{\LT@cols}{@{}p{\textwidth}@{}}{%
    \small\RaggedRight #1{\textbf{#2. }\ignorespaces#3}\par\vspace{6pt}}}
\makeatother

\section{Quantum physics courses at the University of Waterloo}\label{app:courses}
This selected historical catalog was compiled in September 2023 from the university's online course catalog and departmental course pages. The corresponding archived academic calendars provide source context~\cite{waterloo2023undergraduate,waterloo2023graduate}; the selection is not a complete inventory of quantum-related courses. Course entries with identical titles are grouped within each educational level, with all course codes retained. Semicolons separate listings combined on this basis, while slashes preserve the original code groupings. These title-based groups do not imply official cross-listing, and undergraduate and graduate entries remain separate even when their titles match. The catalog contains 13 undergraduate title groups on quantum physics, eight undergraduate title groups covering quantum physics, and 33 graduate title groups.
\Needspace{7\baselineskip}
\begin{longtable}{@{}>{\RaggedRight\arraybackslash}p{\AppendixCodeWidth}>{\RaggedRight\arraybackslash}p{\AppendixCourseWidth}@{}}
\caption{Quantum physics courses at the University of Waterloo.}\label{table:courses}\\
\toprule
\textbf{Course code(s)} & \textbf{Course name}\\
\midrule
\endfirsthead
\caption[]{Quantum physics courses at the University of Waterloo. (Continued.)}\\
\toprule
\textbf{Course code(s)} & \textbf{Course name}\\
\midrule
\endhead
\midrule
\multicolumn{2}{r@{}}{\footnotesize\itshape Continued on next page.}\\
\endfoot
\bottomrule
\endlastfoot
\AppendixGroup{Undergraduate courses on quantum physics}
AMATH~373 & Quantum Theory 1 \\[2.5pt]
AMATH~474;\allowbreak{} PHYS~484 & Quantum Theory 3: Quantum Information and Foundations \\[2.5pt]
CHEM~356 & Introductory Quantum Mechanics \\[2.5pt]
ECE~405;\allowbreak{} PHYS~233 & Introduction to Quantum Mechanics \\[2.5pt]
NE~332 & Quantum Mechanics \\[2.5pt]
PHIL~252 & Quantum Mechanics for Everyone \\[2.5pt]
PHYS~234 & Quantum Physics 1 \\[2.5pt]
PHYS~334 & Quantum Physics 2 \\[2.5pt]
PHYS~434 & Quantum Physics 3 \\[2.5pt]
PHYS~454/\allowbreak{}AMATH~473 & Quantum Theory 2 \\[2.5pt]
PHYS~467/\allowbreak{}CS~467/\allowbreak{}CO~481 & Introduction to Quantum Information Processing \\[2.5pt]
PHYS~468 & Introduction to the Implementation of Quantum Information Processing \\[2.5pt]
PMATH~343 & Introduction to the Mathematics of Quantum Information \\[2.5pt]
\AppendixGroup{Undergraduate courses covering quantum physics}
CHEM~209 & Introductory Spectroscopy and Structure \\[2.5pt]
CHEM~350L & Physical Chemistry Laboratory 2 \\[2.5pt]
PHIL~459 & Studies in the Philosophy of Physics \\[2.5pt]
PHYS~124 & Modern Physics \\[2.5pt]
PHYS~349 & Advanced Computational Physics \\[2.5pt]
PHYS~359 & Statistical Mechanics \\[2.5pt]
PHYS~435 & Current Topics in Condensed Matter Physics \\[2.5pt]
PHYS~444 & Introduction to Particle Physics \\[2.5pt]
\AppendixGroup{Graduate catalog entries}
AMATH~673 & Quantum Theory 2 \\[2.5pt]
AMATH~674 & Quantum Theory 3: Quantum Information and Foundations \\[2.5pt]
AMATH~876/\allowbreak{}QIC~845 & Open Quantum Systems \\[2.5pt]
AMATH~877 & Foundations of Quantum Theory \\[2.5pt]
CHEM~746 & Quantum Chemistry \\[2.5pt]
CO~781;\allowbreak{} QIC~890 & Topics in Quantum Information \\[2.5pt]
CS~766;\allowbreak{} QIC~820 & Theory of Quantum Information \\[2.5pt]
\pagebreak
\AppendixGroup{Graduate catalog entries (continued)}
CS~867 & Advanced Topics in Quantum Computing \\[2.5pt]
ECE~677/\allowbreak{}QIC~885 & Quantum Electronics and Photonics \\[2.5pt]
NANO~707 & From Atoms to Crystals, Quantum Wells, Wires and Dots \\[2.5pt]
PHYS~601 & Perimeter Scholars International Quantum Field Theory 1 \\[2.5pt]
PHYS~603 & Perimeter Scholars International Quantum Field Theory 2 \\[2.5pt]
PHYS~605 & Perimeter Scholars International Quantum Theory \\[2.5pt]
PHYS~635 & Perimeter Scholars International Quantum Information Review \\[2.5pt]
PHYS~638 & Perimeter Scholars International Quantum Gravity \\[2.5pt]
PHYS~639 & Perimeter Scholars International Foundations of Quantum Theory \\[2.5pt]
PHYS~641 & Perimeter Scholars International Explorations in Quantum Information \\[2.5pt]
PHYS~644 & Perimeter Scholars International Explorations in Quantum Gravity \\[2.5pt]
PHYS~645 & Perimeter Scholars International Explorations in Foundations of Quantum Theory \\[2.5pt]
PHYS~701 & Quantum Mechanics 1 \\[2.5pt]
PHYS~702 & Quantum Mechanics 2 \\[2.5pt]
PHYS~703/\allowbreak{}AMATH~873 & Introduction to Quantum Field Theory \\[2.5pt]
PHYS~739 & Quantum Many Body Physics \\[2.5pt]
PHYS~760/\allowbreak{}QIC~860 & Laboratory on Control of Quantum Technology \\[2.5pt]
PHYS~761/\allowbreak{}QIC~861 & Laboratory on Photonic Quantum Technology \\[2.5pt]
PHYS~762/\allowbreak{}QIC~862 & Laboratory on Low Temperature Quantum Technology and Nanofabrication \\[2.5pt]
PHYS~763;\allowbreak{} QIC~863 & Independent Project in Quantum Technology \\[2.5pt]
PHYS~768;\allowbreak{} PHYS~769 & Special Topics in Quantum Information Processing \\[2.5pt]
PHYS~785/\allowbreak{}AMATH~872 & Introduction to Quantum Field Theory for Cosmology \\[2.5pt]
QIC~710/\allowbreak{}PMATH~871/\allowbreak{}PHYS~767/\allowbreak{}CS~768/\allowbreak{}CO~681/\allowbreak{}AMATH~871 & Quantum Information Processing \\[2.5pt]
QIC~750/\allowbreak{}ECE~676 & Quantum Information Processing Devices \\[2.5pt]
QIC~823 & Quantum Algorithms \\[2.5pt]
QIC~880 & Nanoelectronics for Quantum Information Processing \\[2.5pt]
\end{longtable}

\clearpage
\section{Database of works}\label{add:database}

The database contains $122$ distinct works: $115$ journal articles, five conference/proceedings papers, and two book chapters. These totals count publications rather than independent empirical studies. Section~\ref{sec:search} explains the exclusions and scope limitations. Cross-disciplinary background examples retained for refutation-text design and physical-model interpretation are marked with a dagger and explained below their tables. Contextual citations added during revision, including the later instructor survey, are not additional entries in this database.

\Needspace{7\baselineskip}
\begin{longtable}{@{}>{\RaggedRight\arraybackslash}p{\AppendixRefWidth}>{\RaggedRight\arraybackslash}p{\AppendixSummaryWidth}@{}}
\caption{Sources of misconceptions.}\label{table:sources}\\
\toprule
\textbf{Ref.} & \textbf{Summary}\\
\midrule
\endfirsthead
\caption[]{Sources of misconceptions. (Continued.)}\\
\toprule
\textbf{Ref.} & \textbf{Summary}\\
\midrule
\endhead
\midrule
\multicolumn{2}{r@{}}{\footnotesize\itshape Continued on next page.}\\
\endfoot
\bottomrule
\endlastfoot
\AppendixGroup{Sources of knowledge}
\cite{WOS:000340479200002} & Analyzes definitions in The Feynman Lectures, proposing a framework addressing precision, circularity, context, and completeness. \\[0.5pt]
\cite{WOS:000367386000001} & Investigates how different approaches to drawing free body diagrams affect students’ understanding of Newton’s laws. \\[0.5pt]
\cite{WOS:001008829400001} & Identifies potentially misleading optics-textbook representations of image formation and observation. \\[0.5pt]
\cite{WOS:000399912700003} & Examines semiconductor misconceptions and their associations with prior knowledge and information from simulations. \\[0.5pt]
\AppendixGroup{Students}
\cite{WOS:000088224100002} & Examines how language confusion can be mistaken for misconceptions, showing that question misinterpretation can lead to false diagnoses. \\[0.5pt]
\cite{WOS:000749487700001} & Investigates how students with misconceptions focus on misleading information during physics problem-solving. \\[0.5pt]
\cite{WOS:000943130300001} & Investigates how inhibitory control affects the ability to overcome misconceptions about position-velocity relationships. \\[0.5pt]
\cite{WOS:000687380800003} & Examines adolescents’ force understanding, finding an association with working memory but not inhibition or set-shifting. \\[0.5pt]
\cite{WOS:000265992600004} & Investigates five physics graduates’ avoidance of mental simulations, which they regarded as illegitimate tools for reasoning about physics problems. \\[0.5pt]
\cite{WOS:000533707200001} & Reviews physics misconceptions, proposing shared difficulties in reconciling randomness with prediction and physical laws. \\[0.5pt]
\AppendixGroup{Teaching methods}
\cite{cook2006visual} & Examines how visual representations impact cognitive load and learning, emphasizing the role of prior knowledge. \\[0.5pt]
\cite{WOS:000076967200002} & Analyzes conceptual change in kinematics, challenging concept–misconception and expert–novice dichotomies and emphasizing incremental knowledge development. \\[0.5pt]
\end{longtable}

\clearpage
\Needspace{7\baselineskip}
\begin{longtable}{@{}>{\RaggedRight\arraybackslash}p{\AppendixRefWidth}>{\RaggedRight\arraybackslash}p{\AppendixSummaryWidth}@{}}
\caption{Diagnostic methods.}\label{table:diagnostic}\\
\toprule
\textbf{Ref.} & \textbf{Summary}\\
\midrule
\endfirsthead
\caption[]{Diagnostic methods. (Continued.)}\\
\toprule
\textbf{Ref.} & \textbf{Summary}\\
\midrule
\endhead
\midrule
\multicolumn{2}{r@{}}{\footnotesize\itshape Continued on next page.}\\
\endfoot
\bottomrule
\endlastfoot
\AppendixGroup{Tests, inventories, and confidence measures}
\cite{WOS:000269662300015} & Develops a diagnostic test to assess students' misconceptions in quantum physics. \\[0.5pt]
\cite{WOS:000285629600004} & Develops and applies a three-tier test to assess misconceptions about heat and temperature across different education levels. \\[0.5pt]
\cite{WOS:000298904800024} & Develops a three-tier test to diagnose misconceptions about velocity and force in circular motion. \\[0.5pt]
\cite{WOS:000305848300002} & Develops a concept inventory to assess misconceptions about heat, energy, and thermal radiation, showing misconceptions are highly persistent. \\[0.5pt]
\cite{WOS:000318658000001} & Develops and analyzes a relativity concept inventory, examining misconceptions and confidence in answers. \\[0.5pt]
\cite{WOS:000354532600005} & Develops and validates a three-tier diagnostic test for misconceptions about heat, temperature, and internal energy. \\[0.5pt]
\cite{WOS:000373840100003} & Develops and implements a three-tier diagnostic test to assess misconceptions about the photoelectric effect. \\[0.5pt]
\cite{WOS:000400179000007} & Develops and validates a four-tier test to identify misconceptions in geometrical optics. \\[0.5pt]
\cite{WOS:000705678200001} & Uses a four-tier test to identify liquid-pressure misconceptions and distinguish them from knowledge gaps among science teacher candidates. \\[0.5pt]
\cite{WOS:000741862900001} & Develops and validates a three-tier test to assess misconceptions about work, power, and energy, revealing common alternative conceptions. \\[0.5pt]
\cite{fadllan2019analysis} & Identifies student misconceptions in mechanics using a three-tier diagnostic test and clinical interviews. \\[0.5pt]
\cite{diani2019four} & Develops and validates a four-tier diagnostic test to identify student misconceptions in fluid mechanics. \\[0.5pt]
\cite{yuberti2020four} & Develops and evaluates a four-tier diagnostic test with confidence ratings to identify misconceptions about work and simple machines. \\[0.5pt]
\cite{saglam2015confidence} & Examines prospective teachers’ confidence–accuracy calibration on potential-difference questions about parallel circuits. \\[0.5pt]
\cite{WOS:000245278400006} & Develops two-tier tests across physics topics, finding age-related differences but little improvement with age in some concepts. \\[0.5pt]
\cite{WOS:000357865000001} & Develops and tests clicker questions probing students’ interpretation of astronomical photographs and spectra. \\[0.5pt]
\cite{WOS:000297176100001} & Develops an instrument to assess students' understanding of the relationships between force, velocity, and acceleration. \\[0.5pt]
\AppendixGroup{Interviews and open-ended questions}
\cite{WOS:000283554300001} & Uses multi-dimensional cognitive analysis to categorize students’ understanding of heat conduction and identify naive ideas. \\[0.5pt]
\cite{WOS:000485116000001} & Uses visual representations and interviews to diagnose students' preconceptions about friction. \\[0.5pt]
\AppendixGroup{Other diagnostic approaches}
\cite{WOS:000744253000025} & Uses Rasch measurement to analyze misconceptions across multiple science disciplines and identify item difficulty patterns. \\[0.5pt]
\cite{WOS:000568612900004} & Focuses on methods for analyzing student misconceptions using TIMSS data. \\[0.5pt]
\cite{WOS:000766923500011} & Develops a system for diagnosing students' cognitive states and preconceptions to detect misconceptions. \\[0.5pt]
\cite{WOS:001163538600002} & Uses alluvial diagrams and subquestions to analyze depth of conceptual understanding in Newtonian mechanics. \\[0.5pt]
\cite{WOS:000309458400001} & Uses cluster analysis to identify patterns in students’ misconceptions about force and Newton’s third law. \\[0.5pt]
\cite{WOS:000608678900001} & Uses module analysis to identify patterns in misconceptions about electricity and magnetism. \\[0.5pt]
\cite{WOS:000885770300001} & Uses module analysis to identify and compare persistent misconceptions in Newtonian mechanics across different institutions. \\[0.5pt]
\end{longtable}
\AppendixTableNote{Entries are grouped by the main contribution emphasized in this review to help instructors locate relevant approaches. These are not mutually exclusive methods: interviews, for example, may be used to develop tests and inventories. Other diagnostic approaches include Rasch, alluvial, cluster, and module analyses, methodological overviews, and computational diagnosis.}

\clearpage
\Needspace{7\baselineskip}
\begin{longtable}{@{}>{\RaggedRight\arraybackslash}p{\AppendixRefWidth}>{\RaggedRight\arraybackslash}p{\AppendixSummaryWidth}@{}}
\caption{Remediation strategies.}\label{table:remediation}\label{table:remediation1}\label{table:remediation2}\\
\toprule
\textbf{Ref.} & \textbf{Summary}\\
\midrule
\endfirsthead
\caption[]{Remediation strategies. (Continued.)}\\
\toprule
\textbf{Ref.} & \textbf{Summary}\\
\midrule
\endhead
\midrule
\multicolumn{2}{r@{}}{\footnotesize\itshape Continued on next page.}\\
\endfoot
\bottomrule
\endlastfoot
\AppendixGroup{Discussion-based}
\cite{WOS:000179612600005} & Evaluates conceptual assignments and discussions as methods to improve understanding and reduce misconceptions about force and motion. \\[0.5pt]
\cite{WOS:000253202500001} & Develops and evaluates a technology-enhanced learning environment to address misconceptions about seasons. \\[0.5pt]
\cite{WOS:000279714800003} & Uses a bridging analogy-based instruction strategy to address misconceptions about Newton’s Third Law while considering gender and group effects. \\[0.5pt]
\cite{WOS:000287470500024} & Uses constructivist worksheets to successfully address misconceptions about force and motion. \\[0.5pt]
\cite{WOS:000304504800049} & Compares problem-based learning with traditional methods, finding that problem-based learning is more effective at reducing misconceptions. \\[0.5pt]
\cite{WOS:000329795900010} & Examines the effectiveness of the 7E model in addressing misconceptions in electricity compared to traditional methods. \\[0.5pt]
\cite{WOS:000334106800003} & Compares traditional instruction with the 4MAT teaching method, showing it is more effective in reducing misconceptions about electricity. \\[0.5pt]
\cite{WOS:000355004600008} & Investigates concept cartoons embedded in context-based chemistry instruction to address eighth-grade students' alternative conceptions about chemical bonding. \\[0.5pt]
\cite{WOS:000394440400004} & Investigates concept cartoon-embedded worksheets for ninth-grade students' conceptual understanding of Newton's laws of motion. \\[0.5pt]
\cite{leinonen2013overcoming} & Uses hints and peer discussion in a lecture-based intervention to help students overcome misconceptions in thermal physics. \\[0.5pt]
\cite{muller2007tackling} & Uses multimedia to highlight and correct misconceptions in physics, helping students recognize inconsistencies in their reasoning. \\[0.5pt]
\cite{wendt2014effect} & Examines how online collaborative learning affects middle school students' science misconceptions, finding increased misconceptions. \\[0.5pt]
\AppendixGroup{Experiment-based}
\cite{WOS:000345549500012} & Compares lab-based and computer-assisted instruction, finding computer-assisted education more effective in addressing misconceptions about force and motion. \\[0.5pt]
\cite{WOS:001015331200003} & Develops interactive multimedia to address misconceptions about dynamic fluids through cognitive conflict-based learning. \\[0.5pt]
\cite{myneni2013interactive} & Develops and evaluates a virtual physics system that customizes tutoring to address misconceptions in mechanics and energy concepts. \\[0.5pt]
\cite{schneps2014conceptualizing} & Uses virtual 3D solar system simulations on tablets to improve students' understanding of astronomical scale and correct misconceptions. \\[0.5pt]
\cite{hockicko2014correcting} & Uses video analysis and interactive software to correct student misconceptions about car braking distances. \\[0.5pt]
\cite{dutt2012decisions} & Uses dynamic simulations to reduce misconceptions about CO2 stabilization, showing improved understanding through experience-based learning. \\[0.5pt]
\cite{phanphech2019explaining} & Compares circuit-board and virtual-simulation activities, finding learning gains but substantial remaining misconceptions. \\[0.5pt]
\cite{kozhevnikov2013learning} & Compares immersive and non-immersive virtual environments for teaching relative motion, finding that immersion enhances learning of 2D motion. \\[0.5pt]
\cite{korganci2015importance} & Compares traditional, mental modeling, and analogy-based teaching methods for improving student understanding of electric circuits. \\[0.5pt]
\cite{falloon2019using} & Examines the effectiveness of simulations for teaching electricity concepts to young students, highlighting benefits and potential misconceptions. \\[0.5pt]
\cite{WOS:000322006100004} & Compares the effectiveness of cognitive perturbation vs. cognitive conflict in facilitating conceptual change in electricity and magnetism. \\[0.5pt]
\cite{WOS:000751558100012} & Examines the effectiveness of the ICI model with simulations in addressing misconceptions about force and vibration. \\[0.5pt]
\AppendixGroup{Other: instructional model}
\cite{WOS:000281064200006} & Explores a field model for teaching direct current, reporting promising preliminary findings from short high-school pilot trials. \\[0.5pt]
\pagebreak
\AppendixGroup{Reflection-based}
\cite{WOS:000297197200004} & Compares conceptual change methods (texts, analogies, computer animations) and finds that a combination approach is most effective. \\[0.5pt]
\cite{WOS:000318278800002} & Identifies misconceptions about color and tests concept mapping as a strategy to address them, showing long-term learning benefits. \\[0.5pt]
\cite{WOS:000469503400002} & Explores how misconceptions about gravitational acceleration can be leveraged to enhance understanding. \\[0.5pt]
\cite{WOS:000856603000002} & Evaluates conceptual-change texts with cartoons, finding improved electricity understanding but no significant advantage over traditional instruction in attitudes. \\[0.5pt]
\cite{WOS:001109203600004} & Addresses a misconception in relativity and proposes using the twin paradox as a corrective instructional strategy. \\[0.5pt]
\cite{WOS:A1992GX44200002} & Investigates how analogies and examples can be used effectively to correct misconceptions, emphasizing explicit analogy development. \\[0.5pt]
\cite{csahin2010computer} & Develops and introduces a computer-supported conceptual change text to address misconceptions in fluid pressure. \\[0.5pt]
\cite{baser2007effectiveness} & Compares conceptual change instruction with traditional methods for improving students’ understanding of heat and temperature. \\[0.5pt]
\cite{durmucs2010effects} & Evaluates the effectiveness of conceptual change texts and experiments in addressing misconceptions about matter and change. \\[0.5pt]
\cite{ogan2007effects} & Examines how model-based teaching affects pre-service teachers' mental models of lunar phenomena. \\[0.5pt]
\cite{franco2012examining} & Investigates how epistemic beliefs and knowledge representations affect cognitive processing and conceptual change in physics. \\[0.5pt]
\cite{miller2013role} & Analyzes how students' prior knowledge influences their perception and retention of physics lecture demonstrations. \\[0.5pt]
\cite{hein1999using} & Explores a writing-based strategy (folder activity) to help students identify and address physics misconceptions. \\[0.5pt]
\cite{WOS:000184928700003} & Studies how refutational texts can induce conceptual change, helping to correct misconceptions. \\[0.5pt]
\cite{WOS:000442716200001}\textsuperscript{\ensuremath{\dagger}} & Extends the concept of refutation texts beyond science education, showing benefits in history learning. \\[0.5pt]
\cite{WOS:000483008600010} & Examines the effectiveness of explicit vs. implied refutations in science texts, showing that even implied refutations help reduce misconceptions. \\[0.5pt]
\cite{WOS:000642957600001} & Compares different online learning tools and their effectiveness in reducing misconceptions about free-fall motion. \\[0.5pt]
\cite{WOS:A1997XU39100003} & Explores how refutational text structure influences students' conceptual change regarding counterintuitive physics concepts. \\[0.5pt]
\end{longtable}
\AppendixTableNote{\textsuperscript{\ensuremath{\dagger}}Cross-disciplinary background: the history study illustrates refutation-text design, not demonstrated effectiveness in physics instruction.}

\Needspace{7\baselineskip}
\begin{longtable}{@{}>{\RaggedRight\arraybackslash}p{\AppendixRefWidth}>{\RaggedRight\arraybackslash}p{\AppendixSummaryWidth}@{}}
\caption{Examples of misconceptions.}\label{table:examples}\\
\toprule
\textbf{Ref.} & \textbf{Summary}\\
\midrule
\endfirsthead
\caption[]{Examples of misconceptions. (Continued.)}\\
\toprule
\textbf{Ref.} & \textbf{Summary}\\
\midrule
\endhead
\midrule
\multicolumn{2}{r@{}}{\footnotesize\itshape Continued on next page.}\\
\endfoot
\bottomrule
\endlastfoot
\cite{WOS:000574066100002} & Categorizes common misconceptions about sound propagation and acoustic devices. \\[0.5pt]
\cite{WOS:000742147000003} & Conducts a literature review to classify and list common misconceptions about solid friction. \\[0.5pt]
\cite{WOS:000184915100005} & Identifies persistent Aristotelian misconceptions about motion and force in secondary students. \\[0.5pt]
\cite{WOS:000075852700007} & Identifies a misconception that nuclear attraction is evenly shared among electrons, highlighting issues at the physics-chemistry boundary. \\[0.5pt]
\cite{WOS:000269975100007} & Identifies misconceptions about inertia, gravity, and gravitational force among pre-service teachers. \\[0.5pt]
\cite{WOS:000291686800022} & Experimentally supports a corrected photoelectric stopping-potential equation incorporating collector work function and electrode contact potential. \\[0.5pt]
\cite{WOS:000304504700042} & Identifies 32 misconceptions about radiation and radioactivity, comparing students from different disciplines. \\[0.5pt]
\cite{liu2016student} & Reviews 38 force and 15 acceleration misconceptions, grouping reported explanations into four categories. \\[0.5pt]
\cite{WOS:000384323500003} & Identifies alternative conceptions about kinematics, particularly related to the meaning of zero in displacement, speed, velocity, and acceleration. \\[0.5pt]
\cite{WOS:000466445500001} & Identifies persistent alternative worldviews in students’ Force Concept Inventory responses before and after instruction. \\[0.5pt]
\cite{WOS:000568612900006} & Uses international assessment data to identify misconceptions, errors, and misunderstandings in physics and mathematics. \\[0.5pt]
\cite{WOS:A1996UH07600007} & Identifies misconceptions about the causes of seasons, with a focus on inconsistencies in their reasoning. \\[0.5pt]
\cite{WOS:A1996WE28100002} & Examines cognitive resistance to correcting misconceptions about air pressure and vacuums, showing how learners modify knowledge to maintain misconceptions. \\[0.5pt]
\cite{WOS:000405956500015} & Identifies misconceptions about electric current, including potential difference, generators, and magnetic fields, with gender differences in understanding. \\[0.5pt]
\end{longtable}

\clearpage
\Needspace{7\baselineskip}
\begin{longtable}{@{}>{\RaggedRight\arraybackslash}p{\AppendixRefWidth}>{\RaggedRight\arraybackslash}p{\AppendixSummaryWidth}@{}}
\caption{Nature of misconceptions.}\label{table:nature}\\
\toprule
\textbf{Ref.} & \textbf{Summary}\\
\midrule
\endfirsthead
\caption[]{Nature of misconceptions. (Continued.)}\\
\toprule
\textbf{Ref.} & \textbf{Summary}\\
\midrule
\endhead
\midrule
\multicolumn{2}{r@{}}{\footnotesize\itshape Continued on next page.}\\
\endfoot
\bottomrule
\endlastfoot
\cite{WOS:000176433100001} & Reviews persistent beliefs that vision involves emissions from the eyes, resembling historical extramission theories, and discusses instructional responses. \\[0.5pt]
\cite{WOS:000181045900006}\textsuperscript{\ensuremath{\dagger}} & Explores how misapplications of Laplace’s law lead to misunderstandings about alveolar inflation in physiology education. \\[0.5pt]
\cite{WOS:000207961100003} & Uses the Force Concept Inventory (FCI) to diagnose misconceptions about force and motion, finding strong misconceptions about impetus and active force. \\[0.5pt]
\cite{WOS:000228024600001} & Examines why misconceptions about emergent processes (e.g., diffusion, electricity, heat, and evolution) are more robust than misconceptions about direct processes. \\[0.5pt]
\cite{WOS:000265896800003} & Discusses difficulties in interpreting quantum mechanics due to mixing classical concepts with quantum ones, leading to misconceptions. \\[0.5pt]
\cite{WOS:000281414000007} & Addresses a common misconception about thermal behavior of materials, explaining the difference between thin and thick slabs when heated. \\[0.5pt]
\cite{WOS:000285629700012} & Uses analogy-based testing to analyze misconceptions in Newton’s Third Law, finding gender-based and class-level differences. \\[0.5pt]
\cite{WOS:000292149900002} & Compares accelerated-motion tasks, finding linear predictions in action and numerical tasks despite more accurate verbal judgments. \\[0.5pt]
\cite{WOS:000311506200004} & Investigates misconceptions about the microscopic and wave nature of sound across different engineering disciplines. \\[0.5pt]
\cite{WOS:000312337600003} & Investigates intuitive misconceptions about velocity changes, focusing on the "changes-take-time" misconception and its causes. \\[0.5pt]
\cite{WOS:000324551800009} & Categorizes students' alternative conceptions based on epistemological and ontological frameworks. \\[0.5pt]
\cite{WOS:000335541700018} & Compares scientific reasoning and knowledge between students with consistent misconceptions and those with scientific conceptions. \\[0.5pt]
\cite{WOS:000355686800001} & Finds that teacher trainees poorly distinguish attribute from measurement, especially misunderstanding attributes, and discusses implications for teaching. \\[0.5pt]
\cite{WOS:000357483800002} & Compares children’s projectile-motion predictions across ages, finding contextual influences in younger children and impetus reasoning in some older children. \\[0.5pt]
\cite{WOS:000403832500001} & Investigates why students avoid answering conceptual questions and whether learning occurs in a response system setting. \\[0.5pt]
\cite{WOS:000423523700001} & Analyzes incorrect answers in the Force Concept Inventory to understand robust non-Newtonian worldviews. \\[0.5pt]
\cite{WOS:000428300400001} & Compares thermal conceptions across grades, finding improved scientific understanding but some alternative conceptions at similar frequencies. \\[0.5pt]
\cite{WOS:000430180900002} & Compares materialistic conceptions of sound across academic levels, finding misconceptions without a consistent grade-related decline. \\[0.5pt]
\cite{WOS:000600006900001} & Compares misconceptions about impetus theory and projectile motion between students with and without autism, showing differences in adoption rates. \\[0.5pt]
\cite{WOS:000608978200006} & Compares misconceptions about electric circuits across different education levels. \\[0.5pt]
\cite{kulgemeyer2023misconceptions} & Compares accurate and misconception-containing physics videos, finding similar perceived understanding but more misconceptions after inaccurate explanations. \\[0.5pt]
\cite{WOS:000836287700001} & Finds no significant correlation between confidence and reasoning consistency across radioactive-decay questions. \\[0.5pt]
\cite{WOS:000970595800025} & Diagnoses misconceptions about force and motion, showing differences based on sex and grade level, with implications for targeted remediation. \\[0.5pt]
\cite{WOS:001040148700001} & Examines misconceptions in scientific reasoning using confidence-based assessment, showing high confidence in incorrect explanations. \\[0.5pt]
\cite{WOS:001088164200001} & Differentiates between misconceptions and a mere lack of knowledge in understanding electric circuits. \\[0.5pt]
\cite{WOS:001129334900001} & Investigates misconceptions about AI trustworthiness in physics learning, showing a high level of misplaced trust. \\[0.5pt]
\cite{WOS:001167417300002} & Surveys astronomy interest, knowledge, and misconceptions, comparing student groups and examining associations with educational background. \\[0.5pt]
\end{longtable}
\AppendixTableNote{\textsuperscript{\ensuremath{\dagger}}Cross-disciplinary background: the physiology example concerns the limits of applying a physical model to a biological system, not misconceptions diagnosed in quantum students.}

\endgroup

\end{document}